\documentclass[%
reprint,
 amsmath,amssymb,
 aps,
]{revtex4-2}
\usepackage[colorinlistoftodos]{todonotes}
\usepackage{comment}
\usepackage{ulem}
\usepackage{hyperref}
\usepackage{graphicx}
\usepackage{dcolumn}
\usepackage{caption}
\usepackage{subcaption}
\usepackage{array}
\newcolumntype{P}[1]{>{\centering\arraybackslash}p{#1}}
\usepackage{bm,color}
\usepackage{amssymb}
\usepackage{amsmath}
\usepackage{comment}
\usepackage{makecell,tabularx,multirow}
\usepackage{mathtools}
\usepackage{times}
\usepackage{bigints}
\usepackage{enumitem}
\usepackage{float}
\usepackage{array}
\usepackage{booktabs}
\usepackage{graphicx}
\usepackage{dcolumn}
\usepackage{bm}
\usepackage{comment}

\begin{document}

\preprint{APS/123-QED}
\title{Kalb-Ramond field induced cosmological bounce in generalized teleparallel gravity}

\author{Krishnanand K. Nair}
 \email{krishnanandknair@gmail.com}
\author{Mathew Thomas Arun}%
 \email{mathewthomas@iisertvm.ac.in}
\affiliation{
School of Physics, Indian Institute of Science Education and Research, Thiruvananthapuram 695551, India\\
}%

\begin{abstract}
One of the important open questions in high-energy physics is to understand the lack of evidence of Kalb-Ramond (KR) field, in particular in the present day cosmology. In this paper we aim to address this issue by showing that a bounce scenario in the evolution of the Universe strongly advocates their elusiveness, even if their energy density was very large to start with. We consider the Kalb-Ramond field and its effects in the context of generalized teleparallel gravity in (3+1) dimensions. Teleparallel gravity is a description of gravitation in which the tetrads are the dynamical degrees of freedom, and the torsion arising from fields with spin are accommodated naturally as field strength tensors. In order to describe the coupling prescription, we address the correct generalization of the Fock-Ivanenko derivative operator for an n-form tensor field. By varying with respect to the tetrads, this  rank-2 field is shown to source the teleparallel equivalent of Einstein's equations. We study the possibility of reproducing two well-known cosmological bounce scenarios, namely, symmetric bounce and matter bounce in four-dimensional spacetime with with the Friedmann-Lemaître-Robertson-Walker metric and observe that the solution requires the KR field energy density to be localized near the bounce. The crucial result in our work is that this feature also naturally explains the lack of cosmological evidence of the rank-2 field in the present day Universe for the matter-bounce scenario. Thus, among the bouncing cosmologies, latter is favored over the former. 
\end{abstract}

\maketitle


\section{Introduction}

The Kalb-Ramond (KR) field has been understood to be essential to correctly reproduce the low-energy string effective action \cite{Schwarz:2000ew,Mukhopadhyaya:2002jn}. These antisymmetric tensor fields constitute the field content of all superstring models and must have significant imprint during the primordial epoch of the evolution of the Universe. Apart from string inspired models, the KR field arise in higher-dimensional theories, that aim to unify gravity and electromagnetism. Still, the KR field is not yet detected in any of the experiments.\cite{Das:2018jey}.

In this paper, we aim to address the KR field in a generalized teleparallel setup and show that a natural explanation for its absence in the present-day Universe is realized in bouncing cosmology. Apart from the benefit that, here, gravity is understood as a gauge theory of the translation group \cite{Aldrovandi:2013wha}, along with a conserved energy-momentum gauge current, this description also naturally accommodates the effects of fields with a spin quantum number through tetrads, which form the dynamical variables, instead of the metric. Generalized teleparallel gravity also naturally accommodates an explanation to cosmological phenomenon like the late-time acceleration of the Universe \cite{Shie:2008ms, Bengochea:2008gz,Ao:2010mg, Wu:2010mn, Linder:2010py, Myrzakulov:2010vz,Cai:2015emx,Wright:2016ayu}.

Here, we consider a generalized teleparallel gravity setup in $(3+1)$ dimensions appended by an action of the Kalb-Ramond field. With the appropriate generalization of the Fock-Ivanenko derivative operator for the KR field, we compute the equivalent of Einstein's equations by varying the action with respect to the tetrads. This gives the equivalent energy-momentum tensor of the antisymmetric field. With the setup in place we now study the requirement to achieve bouncing cosmology. 

Models with bounces \cite{Novello:2008ra, Battefeld:2014uga, Brandenberger:2016vhg} provide an elegant solution to the initial singularity in the big bang paradigm and, in some instances, could generate a scale-invariant power-law spectrum \cite{Brandenberger:2012zb} as well. Even though there have been immense efforts carried out in modified gravity theories with higher-order corrections\cite{Brustein:1997cv, Biswas:2006bs} and in braneworld scenarios\cite{Kehagias:1999vr, Saridakis:2007cf}, it is interesting to understand these phenomena in the teleparallel equivalent of General Relativity (TEGR) \cite{Aldrovandi:2013wha}. In this paper, we explicitly compute the energy spectrum of the tensor field and the appropriate teleparallel gravity model for symmetric and matter bounce scenarios. We show that the energy and pressure densities of the tensor field are indeed localized at $t=0$, which acts as the source for the bounce. We find that, in the case of symmetric bounce, a significant fraction of the energy density of KR field remains to the present day Universe, whereas, in the context of matter bounce, the energy density of the KR field drastically decreases from $3$ M$_{Pl}{}^4$  at the bounce, to $\sim 0$ at $t=t_0$. Hence, we show that the null results from searches for the KR field strongly suggests matter bounce for the cosmic evolution.\\
The paper is categorized as follows. We start with a brief review of TEGR formalism in Sec.\eqref{sec:TEGR} and introduce Kalb-Ramond fields as a source of torsion. In Sec.\eqref{sec:FID}, we explain the minimal coupling prescription and develop the Fock-Ivanenko operator for the Kalb-Ramond field. As an application to cosmology, in Sec.\eqref{sec:bounces}, we compute the energy density and pressure density of the KR fields in the generalized teleparallel setup that will lead to correct expansion coefficients in symmetric and matter bounce scenarios. Finally, in Sec.\eqref{sec:discussion} we summarise our results.

\section{Teleparallel Equivalent of General Relativity}
\label{sec:TEGR}
In Einstein's General Relativity (GR), the affine connection is taken to be  torsionless and satisfies the metricity condition,
\begin{equation}
    \nabla_\mu g_{\nu \rho}=0
    \label{nmcond}
\end{equation}
 where $\nabla_\mu$ is the covariant derivative with the Levi-Civita $\Tilde{\Gamma}^{\mu}_{\nu \rho}$ playing the role of affine connection. 
 However, in teleparallel gravity (TG) , the Levi-Civita affine connection is replaced by the Weitzenb$\ddot{\text{o}}$ck connection, which is torsionfull but curvatureless and satisfies the metricity condition Eq.\eqref{nmcond}. \\
 
Although teleparallel gravity is an alternative to General Relativity, they are conceptually distinct \cite{Bahamonde:2021gfp}. 
In TEGR, the spacetime metric is constructed out of tetrads ($h^a{}_\mu$), which are the dynamical degrees of freedom, as,
\begin{equation}
   g_{\mu \nu}= \eta_{ab}h^a{}_\mu h^b{}_\mu \ ,
    \label{mp}
\end{equation}
where $\eta_{a b}$ is the Minkowski metric of the tangent space. The tetrads $h^a{}_\mu$ could be written in terms of flat-space tetrads ($e^a{}_\mu $ = $\partial_\mu x^a$) as \cite{Bahamonde:2021gfp}
\begin{equation}
 h^a{}_\mu =  e^a{}_\mu + \omega^a{}_{b\mu}x^b +  A^a{}_\mu \ .
     \label{cp1}
\end{equation}
The flat-space tetrads satisfy the relation $\eta_{\mu\nu} = \eta_{ab} e^a_\mu e^b_\nu$, where $\eta_{\mu \nu}$ is the metric of Minkowski spacetime and the spin connection ($\omega^a{}_{b\mu}$) is given by $\omega^a{}_{b\mu}=\Lambda^a{}_c\partial_\mu\Lambda_b{}^c$, where $\Lambda$ is the Lorentz matrix and the translational connection on the tangent space is denoted by $A^a{}_\mu$. Note that, in this paper, we will be referring Greek indices ($\mu$, $\nu$) to the spacetime manifold and the Latin indices (a, b) to the local Minkowski tangent space.
We also assume the signature of  $\eta_{ab}$ as  diag $ (- + + + )$.\\
Now, the Weitzenb$\ddot{\text{o}}$ck connection $\Gamma^{\rho}{}_{\mu \nu}$ can be written as \cite{wb}
\begin{equation}
\Gamma^{\rho}{}_{\mu \nu}=h_{a}{}^{\rho} \partial_{\mu} h^{a}{ }_{\nu} + h_{a}{}^{\rho}\omega^a{}_{b\mu}h^{b}_{\nu} \ .
\label{wbconn}
\end{equation}
 Since we are interested in the evolution of the Universe, we stick to a particular choice of tetrads given in Eq.eq\ref{tetrad} corresponding to the flat Friedmann-Lemaître-Robertson-Walker (FLRW) spacetime for which the spin connection $\omega^a{}_{b\mu}=0$ \cite{Gonzalez:2011dr,Krssak:2015oua, Farrugia:2018gyz, Capozziello:2018qcp, Bahamonde:2016grb}. Given this solution, one can easily show that the Weitzenb$\ddot{\text{o}}$ck covariant derivative of the tetrads vanish identically, thus satisfying the metricity condition,
\begin{equation}
\nabla_{\mu} h_{\nu}{}^{A} \equiv \partial_{\mu} h^{A}{ }_{\nu}-\Gamma_{\mu \nu}^{\rho} h_{\rho}{}^{A}=0 \ ,
\label{WD-tetrad}
\end{equation}
where $\nabla_{\mu}$ represents the covariant derivative constructed with the Weitzenb$\ddot{\text{o}}$ck connection.\\
Now, the torsion tensor could be constructed from the Weitzenb$\ddot{\text{o}}$ck connection as given below,
\begin{equation}
T^{\rho}{}_{\mu \nu}=\Gamma^{\rho}{}_{\mu \nu}-\Gamma^{\rho}{}_{\nu\mu } \ .
\label{torsion}
\end{equation}
Using Eq.\eqref{cp1} and Eq.\eqref{wbconn}, it is straightforward to see that the torsion tensor acts as the field strength of the translation potential $A^a{}_\mu$, for spin connection $\omega^a{}_{b\mu}=0$~\cite{Bahamonde:2021gfp}
\begin{equation}
    T^a{}_{\mu \nu}=h^a{}_\rho T^\rho{}_{\mu \nu}=\partial_\mu A^a{}_\nu - \partial_\nu A^a{}_\mu \ .
\end{equation}
The Weitzenb$\ddot{\text{o}}$ck connection in teleparallel gravity  $\Gamma^{\rho}{}_{\mu \nu}$ and the Levi-Civita connections $\Tilde{\Gamma}^{\rho}{}_{\mu \nu}$ in GR are then mathematically related as
\begin{equation}
\Gamma^{\rho}{}_{\mu \nu}-K^{\rho}{}_{\mu \nu} \equiv \Tilde{\Gamma}^{\rho}{}_{\mu \nu} \ ,
\label{rel1}
\end{equation}
where $K^{\rho}{}_{\mu \nu}$ is  the contorsion tensor given by
\begin{equation}
K^{\rho}{}_{\mu \nu}=\frac{1}{2}\left(T_{\mu}{ }^{\rho}{ }_{\nu}+T_{\nu}{ }^{\rho}{ }_{\mu}-T^{\rho}{}_{\mu \nu}\right) \ .
\label{contorsion}
\end{equation}
Note that, we use overtilde to represent quantities calculated using the Levi-Civita connection in GR to distinguish it from teleparallel gravity in this paper. It is straightforward to show that the curvature of the Weitzenb$\ddot{\text{o}}$ck connection also vanishes.
\begin{equation}
{R}^{\rho}{}_{\lambda \mu \nu}(\Gamma) = 0 \ .
\end{equation}
The dual torsion tensor is defined as 
\begin{equation}
S^{\rho \mu \nu}= \frac{1}{2}\left[K^{\mu \nu \rho}-g^{\rho \nu} T^{\lambda \mu}{}_{\lambda}+g^{\rho \mu} T^{\lambda \nu}{}_{ \lambda}\right] \ .
\label{dualtorsion}
\end{equation}
Finally, we define a quadratic function of torsion called the torsion scalar T given by,
\begin{equation}
T= T_{\rho \mu \nu} S^{\rho \mu \nu}=T^{\rho}{}_{\mu \nu} T_{\rho}{ }^{\mu \nu} /2+T^{\rho}{}_{\mu \nu} T^{\nu \mu}{}_{\rho}-2 T^{\rho}{}_{\mu \rho} T^{\nu \mu}{ }_{\nu} \ .
\label{tscalar}
\end{equation}
The gravitational Lagrangian using the torsion scalar can be written as
\begin{equation}
\mathcal{L}_{G}=-\frac{h}{16 \pi G} T \ ,
\end{equation}
where $h = det(h^a_\mu) = \sqrt{-g}$

Using Eq.\eqref{rel1} in the above action and reformulating the above Lagrangian in terms of Levi-Civita connection, we can obtain the mathematical relation between the torsion scalar $T$ in teleparallel gravity and the Ricci scalar $\Tilde{R}$ in GR 
\begin{equation}
    T \equiv -\Tilde{R} + B \ ,
\end{equation}
where $\Tilde{R}$ is the Ricci scalar and $B=2\Tilde{\nabla}_\mu(T^\nu{}_{\nu}{}^{\mu})$ is a total divergence term. Thus this action is equivalent to the Einstein-Hilbert action, which gives Einstein's field equations of gravity \cite{Bahamonde:2015zma}. 
\section{Coupling prescription using Fock–Ivanenko derivative operator in the teleparallel geometry}
\label{sec:FID}
In Minkowski space, the dynamics of the Kalb-Ramond field is described by the Lagrangian \cite{Kalb:1974yc}
\begin{equation}
    \mathcal{L}_{KR} = -H_{abc}H^{abc} \ ,
\end{equation}
where 
\begin{equation}
    H_{abc} = \partial_a B_{bc}+\partial_b B_{ca} + \partial_c B_{ab} \ ,
\end{equation}
is the field strength of the Kalb-Ramond field $B_{ab}$, which is a rank-2 antisymmetric tensor field.\\
On varying the action with respect $B_{ab}$, we get the field equations
\begin{equation}
    \partial_a H^{abc}=0 \ ,
    \label{fe1}
\end{equation}
along with the Bianchi identity
\begin{equation}
    \partial_{[a}H_{bcd]}=0 \ ,
\end{equation}
For the Lorentz gauge $\partial_a B^{ab} = 0$, the field equation Eq.\eqref{fe1} becomes
\begin{equation}
    \partial_c \partial^c B^{ab} = 0 \ .
\end{equation}
However in teleparallel gravity, the existence of torsion destroys the gauge invariance of the theory when the KR field is used as the source of the equation of motion of the field. If we assume the coupling prescription given by
 \begin{equation}
\begin{aligned}
&\eta^{a b} \rightarrow g^{\mu \nu}=\eta^{a b} h_{a}^{\mu} h_{b}^{\nu} \\
&\partial_{a} \rightarrow \nabla_{\mu} \equiv \partial_\mu - \Gamma_\mu \ ,
\end{aligned}
\label{mcouplingtele}
\end{equation}
where $\Gamma$ is the Weitzenb$\ddot{\text{o}}$ck connection, the KR field strength takes the form,
\begin{equation}
\begin{aligned}
    H_{\mu \nu \rho} &= {\nabla}_{\mu}B_{\nu \rho}+{\nabla}_{\rho}B_{\mu \nu}+{\nabla}_{\nu}B_{\rho \mu}\\
    &=3 \partial_{[\mu}B_{\nu \rho]}+3T^\sigma{}_{[\mu \nu}B_{\rho]\sigma} \ .
    \label{nonmcoup}
\end{aligned}
\end{equation}
The last term in Eq.\eqref{nonmcoup} indicates the nonminimal coupling of torsion with the KR field in teleparallel geometry and thus Eq.\eqref{nonmcoup} is not invariant under U(1) gauge transformation.\\

In order to keep the transformation gauge invariant, in the framework of teleparallel geometry,  one needs to use the minimal coupling prescription \cite{deAndrade:1997cj},
\begin{equation}
\begin{aligned}
&\eta^{a b} \rightarrow g^{\mu \nu}=\eta^{a b} h_{a}^{\mu} h_{b}^{\nu} \\
&\partial_{a} \rightarrow \mathcal{D}_{\mu}=\partial_{\mu}-\frac{i}{2}\Omega^{a b}{ }_{\mu} J_{a b} \ ,
\end{aligned}
\label{mcoupling}
\end{equation}
where $\mathcal{D}_{\mu}$ is the Fock–Ivanenko derivative operator \cite{Fock1929}, which acts only on the local Lorentz indices. Here, $\Omega^{a b}{ }_{\mu}$ is given by
\begin{equation}
\Omega^{a b}{ }_{\mu}=-h_{\rho}^{a} K^{\rho \nu}{ }_{\mu} h_{\nu}^{b} \ ,
\label{spinconnection1}
\end{equation}
and $J_{ab}$ is the generator in the appropriate representation of the Lorentz group. For instance, $J_{ab}$ acting on any $n$-form field could be written as
\begin{equation}
\begin{aligned}
J_{ab}{}(B^{i_1i_2...i_n}) = &  i\big(\delta^{i_1}_a\eta_{bc}-\delta^{i_1}_b\eta_{ac}\big)B^{ci_2...i_n}\\&
 +i\big(\delta^{i_2}_a\eta_{bc}-\delta^{i_2}_b\eta_{ac}\big)B^{i_1c...i_n}\\&
  + .... + i\big(\delta^{i_n}_a\eta_{bc}-\delta^{i_1}_b\eta_{ac}\big)B^{i_1i_2...c} \ .
\label{Jab-Bab1}
\end{aligned}
\end{equation}
 It is also important to note that FIDO in teleparallel gravity is equivalent to the Levi-Civita covariant derivative in the Einstein GR in the absence of contorsion, as shown in the Appendix \eqref{apa}. More importantly, with this coupling prescription, torsion does not violate the gauge symmetry of Kalb-Ramond theory. 
 \\
 
Using Eq.\eqref{spinconnection1} and Eq.\eqref{Jab-Bab1}, we get the Fock–Ivanenko derivative acting on $B^{ab}$ as
\begin{equation}
\begin{aligned}
 \mathcal{D}_\mu B^{ab} &= \partial_\mu B^{ab} - \frac{i}{2}\Omega^{c d}{ }_{\mu}\big(i(\delta^a_c\eta_{dg}-\delta^a_d\eta_{cg})\big)B^{gb}\\
    &-\frac{i}{2}\Omega^{c d}{ }_{\mu}\big(i(\delta^b_c\eta_{dg}-\delta^b_d\eta_{cg})\big)B^{ag}\\
   & = \partial_\mu B^{ab} - K^{\rho}{}_{\nu \mu}h_{d}{}^{\nu}h^{[a}{}_{\rho}B^{b]d} \ .
    \label{TFD-Bab}
    \end{aligned}
\end{equation}
Any spacetime tensor $B^{\mu \nu}$ can be transformed to a Lorentz tensor $B^{ab}$ by
\begin{equation}
    B^{ab} =  h^{a}{ }_{\mu} h^{b}{ }_{\nu} B^{\mu \nu} \ .
    \label{spacet-lorentz1}
\end{equation}
Now, using Eq.\eqref{spacet-lorentz1} and making use of Eq.\eqref{WD-tetrad} in Eq.\eqref{TFD-Bab}, we have the teleparallel version of the covariant derivative
\begin{equation}
\mathcal{D}_\mu B^{ab}= h^{a}{ }_{\rho} h^{b}{ }_{\sigma}\stackrel{\circ}{\nabla}_\mu B^{\rho \sigma} \ ,
\end{equation}

with
\begin{equation}
    \stackrel{\circ}{\nabla}_\mu B^{\rho \sigma} = {\nabla}_\mu  B^{\rho \sigma} - K^{\rho}_{\lambda \mu} B^{\lambda \sigma} - K^{\sigma}_{\lambda \mu} B^{\rho \lambda}
\end{equation}
where ${\nabla}_\mu  B^{\rho \sigma}$ is the Weitzenb$\ddot{\text{o}}$ck covariant derivative given as
\begin{equation}
    {\nabla}_\mu  B^{\rho \sigma}=  \partial_\mu  B^{\rho \sigma} +  \Gamma^{\rho}_{\mu \lambda } B^{\lambda \sigma} + \Gamma^{\sigma}_{\mu \lambda } B^{\rho \lambda} \ .
\end{equation}
Thus the teleparallel version of minimal coupling prescription is given as
\begin{equation}
    \partial_{a} \rightarrow \stackrel{\circ}{\nabla}_\mu \equiv \partial_{\mu}+\Gamma_{\mu}-K_{\mu} \ .
\end{equation}
The Fock–Ivanenko derivative $\stackrel{\circ}{\nabla}_\mu$ in  Eq.\eqref{mcoupling} turns out to be the Weitzenb$\ddot{\text{o}}$ck connection in teleparallel gravity minus the contorsion tensor. \\

With the correct prescription ready, let us now consider the Kalb-Ramond action in the teleparallel background as follows,
\begin{equation}
    \mathcal{L}_m = -hH_{\mu \nu \rho}H^{\mu \nu \rho} \ ,
\end{equation}
where $h=\sqrt{-g}$ and  $H_{\mu \nu \rho}$ is given as,
\begin{equation}
\begin{aligned}
    H_{\mu \nu \rho} &= \stackrel{\circ}{\nabla}_{\mu}B_{\nu \rho}+\stackrel{\circ}{\nabla}_{\rho}B_{\mu \nu}+\stackrel{\circ}{\nabla}_{\nu}B_{\rho \mu}\\
    &= \partial_\mu B_{\nu \rho}+\partial_\rho B_{\mu \nu} + \partial_\nu B_{\rho \mu} \ ,
    \end{aligned}
\end{equation}
which is U(1) gauge invariant. The teleparallel version of field equation is given as
\begin{equation}
    \stackrel{\circ}{\nabla}_{\mu}H^{\mu \nu \rho} = 0 \ .
    \end{equation}
And the teleparallel version of the Bianchi identity can be written as
\begin{equation}
    \stackrel{\circ}{\nabla}_{[\mu}H_{\nu \rho \sigma]} = 0 \ .
    \label{telebianchi}
\end{equation}
Assuming Lorentz gauge $\stackrel{\circ}{\nabla}_{\mu}B^{\mu \nu}=0$, and using the commutation relation
\begin{equation}
    \big[\stackrel{\circ}{\nabla}_{\mu},\stackrel{\circ}{\nabla}_{\nu}\big] B^{\lambda\mu}=    -Q^{\lambda}_{\sigma\mu\nu}B^{\sigma\mu}-Q_{\mu \nu}B^{\lambda\mu} \ ,
\end{equation}
where
\begin{equation}
   Q_{\rho \mu \nu}^{\theta}=\nabla_{\mu} K^{\theta}{}_{\rho \nu}-K^{\theta}{}_{\sigma \nu} K^{\sigma}{}_{\rho \mu}-\nabla_{\nu} K^{\theta}{}_{\rho \mu}+K^{\theta}{}_{\sigma \mu} K^{\sigma}{}_{\rho \nu} \ ,
\end{equation}
we can derive the  field equations in teleparallel gravity to be
\begin{equation}
    \stackrel{\circ}{\nabla}_{\mu}\stackrel{\circ}{\nabla}^{\mu}B^{\nu \lambda} - Q^{\nu \lambda \sigma \mu}B_{\sigma \mu}-2Q_{\mu}{}^{[\nu}B^{\lambda]\mu} = 0 \ .
\end{equation}
\section{Non-Singular Cosmological bounce in the presence of Kalb Ramond field}
\label{sec:bounces}
To study the cosmological bouncing in $F(T)$ gravity, lets consider the flat homogeneous isotropic FLRW metric, 
\begin{equation}
    ds^2 = -dt^2+a(t)^2(dx^2+dy^2+dz^2) \ ,
    \label{flrwmetric}
\end{equation}
where $a(t)$ is the scale factor, which is a function of $t$.
Corresponding to this metric, the  tetrads become,
\begin{equation}
h^a{}_\mu = \text{diag}\big(1, a(t),a(t),a(t)\big) \ .
\label{tetrad}
\end{equation}
In this geometry, the nonzero components of the Weitzenb$\ddot{\text{o}}$ck connection Eq.\eqref{wbconn}, torsion tensor Eq.\eqref{torsion}, contorsion tensor Eq.\eqref{contorsion} and dual torsion tensor Eq.\eqref{dualtorsion} can be derived as
\begin{gather}
    \Gamma^i{}_{0i} = H \ ,\\
     T^i{}_{i0} = -T^i{}_{0i} = -H \ ,\\
      K^0{}_{ii} = -Ha(t)^2 \ ,\\
      K^i{}_{0i} = -H \ ,\\
      S^i{}_{0i} = -S^i{}_{i0}=H \ ,
\end{gather}
where $H=\frac{a'(t)}{a(t)}$ is the Hubble parameter.
Thus, we can compute the torsion scalar using Eq.\eqref{tscalar} as
\begin{equation}
    T=6H^2 \ .
\end{equation}
Our objective is to find the functional form of the gravitational Lagrangian $F(T)$ that can give rise to nonsingular bouncing cosmology in the presence of Kalb-Ramond fields in the FLRW geometry. To do this, let us consider the action,
\begin{equation}
    \begin{aligned}
        S = \frac{1}{2\kappa^2}\Bigg[\int d^4x \ h\Big(\operatorname{F(T)}+\Lambda\Big) \Bigg]-\frac{1}{2}\int d^4x \ h\ H_{\mu \nu \rho}H^{\mu \nu \rho}  \ ,
    \end{aligned}
\end{equation}
where $F(T)=-T+f(T)$, $\kappa=\sqrt{8\pi G}$ and $\Lambda$ is the cosmological constant. On varying this action with respect to the tetrads $h^{a}_\mu$ \cite{Bahamonde:2015zma}, we get the following equations of motion,
\begin{equation}
    \begin{aligned}
        M_\mu{}^\nu\equiv & \ 2hf_{TT}\partial_\mu T S_\nu{}^{\mu \lambda} + 2f_Te^a{}_\nu\partial_\mu(hS_a{}^{\mu \lambda})\\ &-2hf_T T^\sigma{}_{\mu \nu} S_\sigma{}^{ \lambda \mu}
        -h\big(f+\Lambda\big)\delta^\lambda{}_\nu\\  = & h  \kappa^2\Big(3H_{\mu \rho \sigma}H^{\nu \rho \sigma}-\frac{1}{2}\delta_\mu{}^\nu H_{\rho \sigma \lambda}H^{\rho \sigma \lambda}\Big) \ ,
    \end{aligned}
    \label{eom1}
\end{equation}
\begin{figure*}
     \centering
     \begin{subfigure}[t!]{0.44\textwidth}
         \includegraphics[scale=0.45]{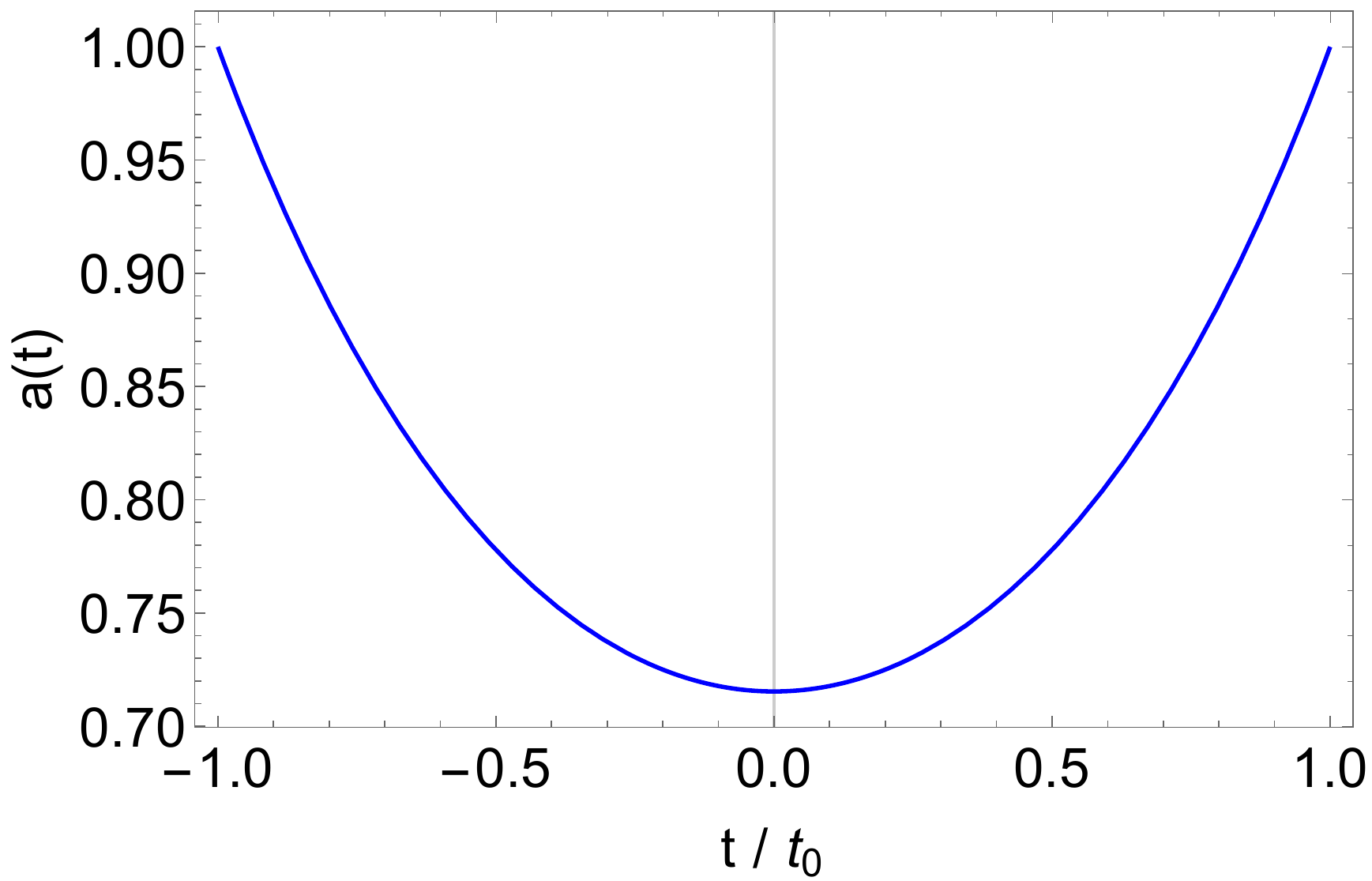}
         \caption{}
         \label{a(t)sb}
     \end{subfigure}
     \hfill
     \begin{subfigure}[t!]{0.44\textwidth}
         \includegraphics[scale=0.45]{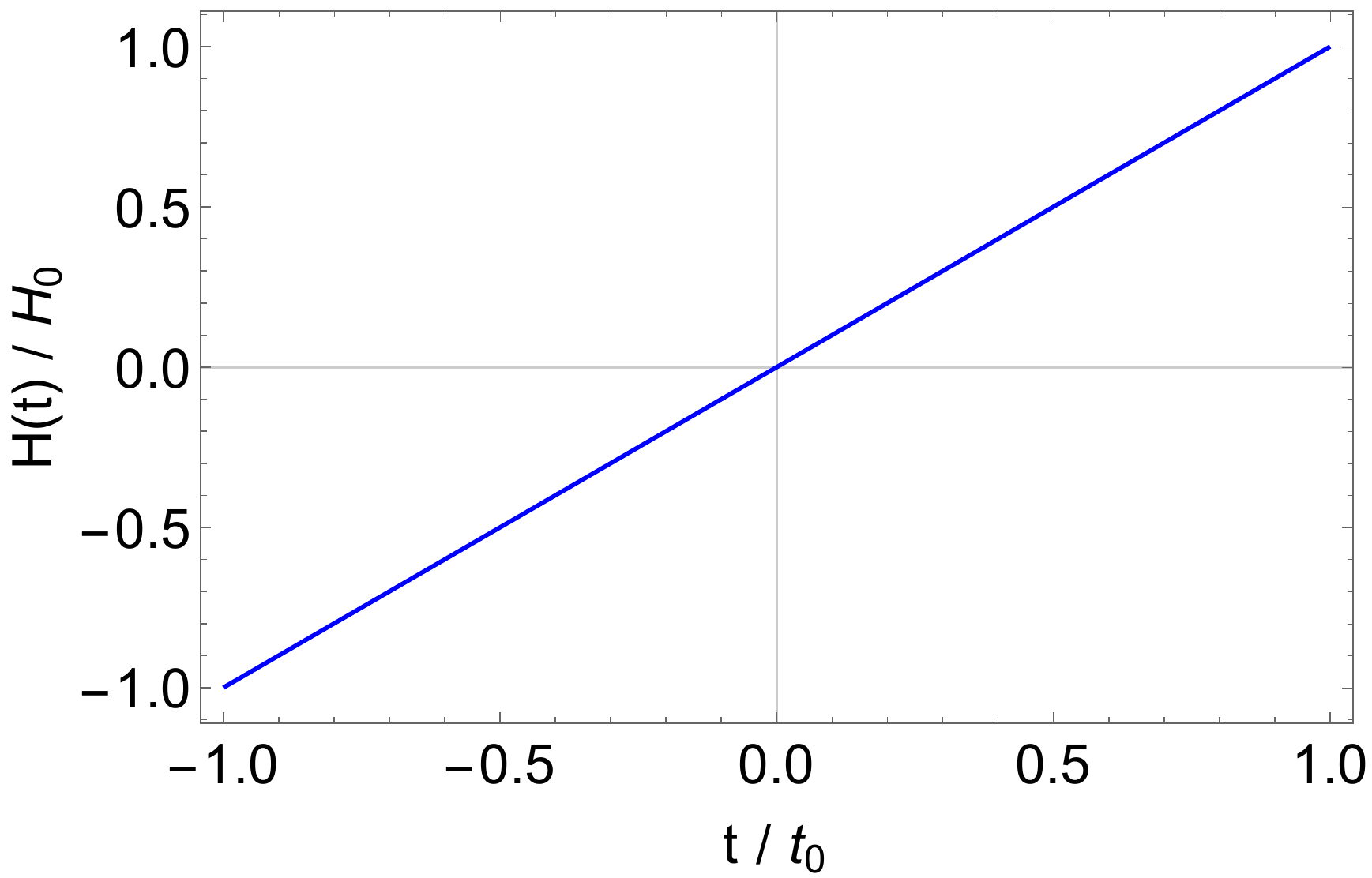}
         \caption{}
         \label{h(t)sb}
     \end{subfigure}
     \newline
     \begin{subfigure}[t!]{0.44\textwidth}
          \includegraphics[scale=0.45]{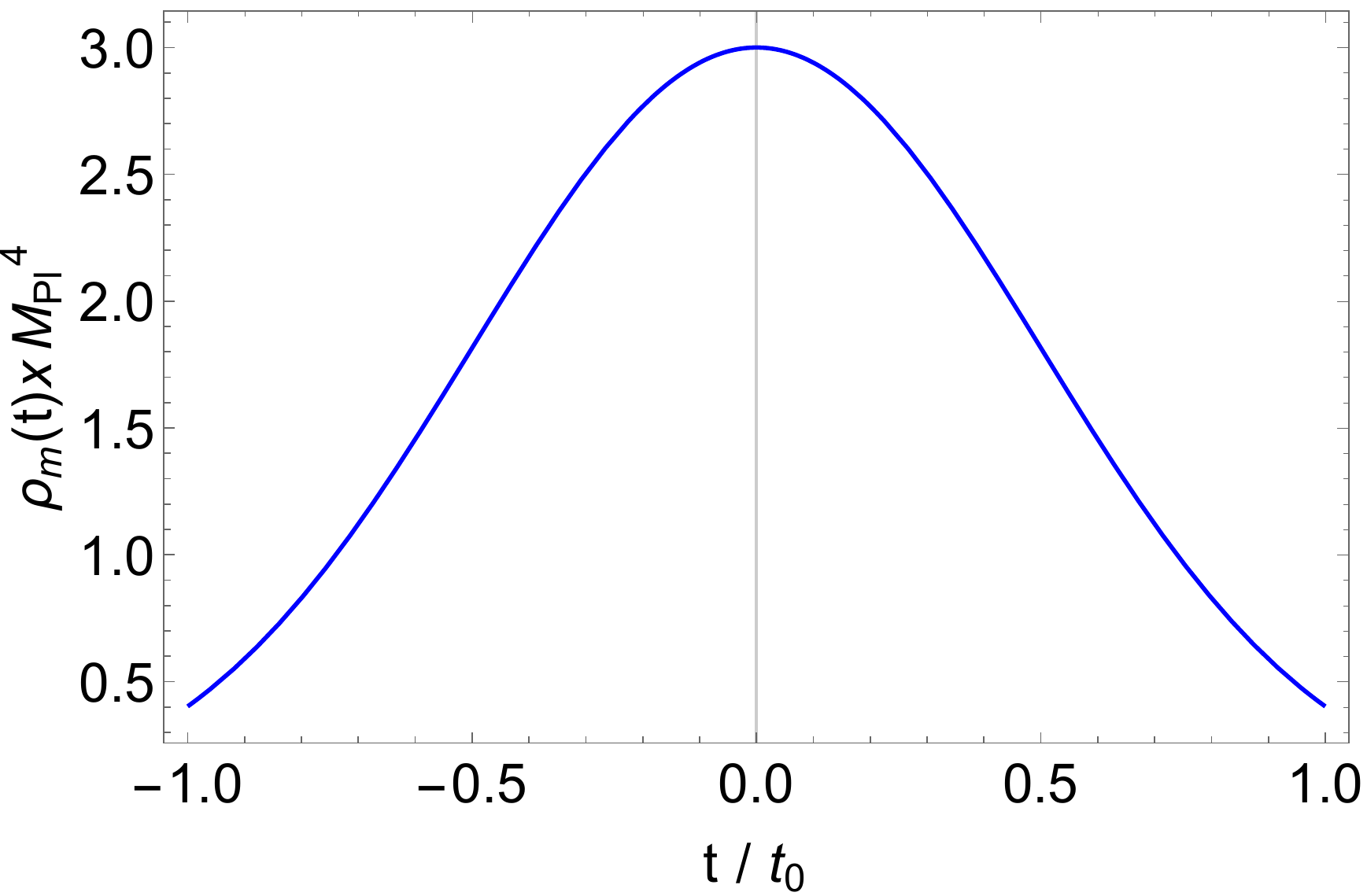}
         \caption{}
         \label{rhosb}
     \end{subfigure}
      \hfill
     \begin{subfigure}[t!]{0.44\textwidth}
         \includegraphics[scale=0.45]{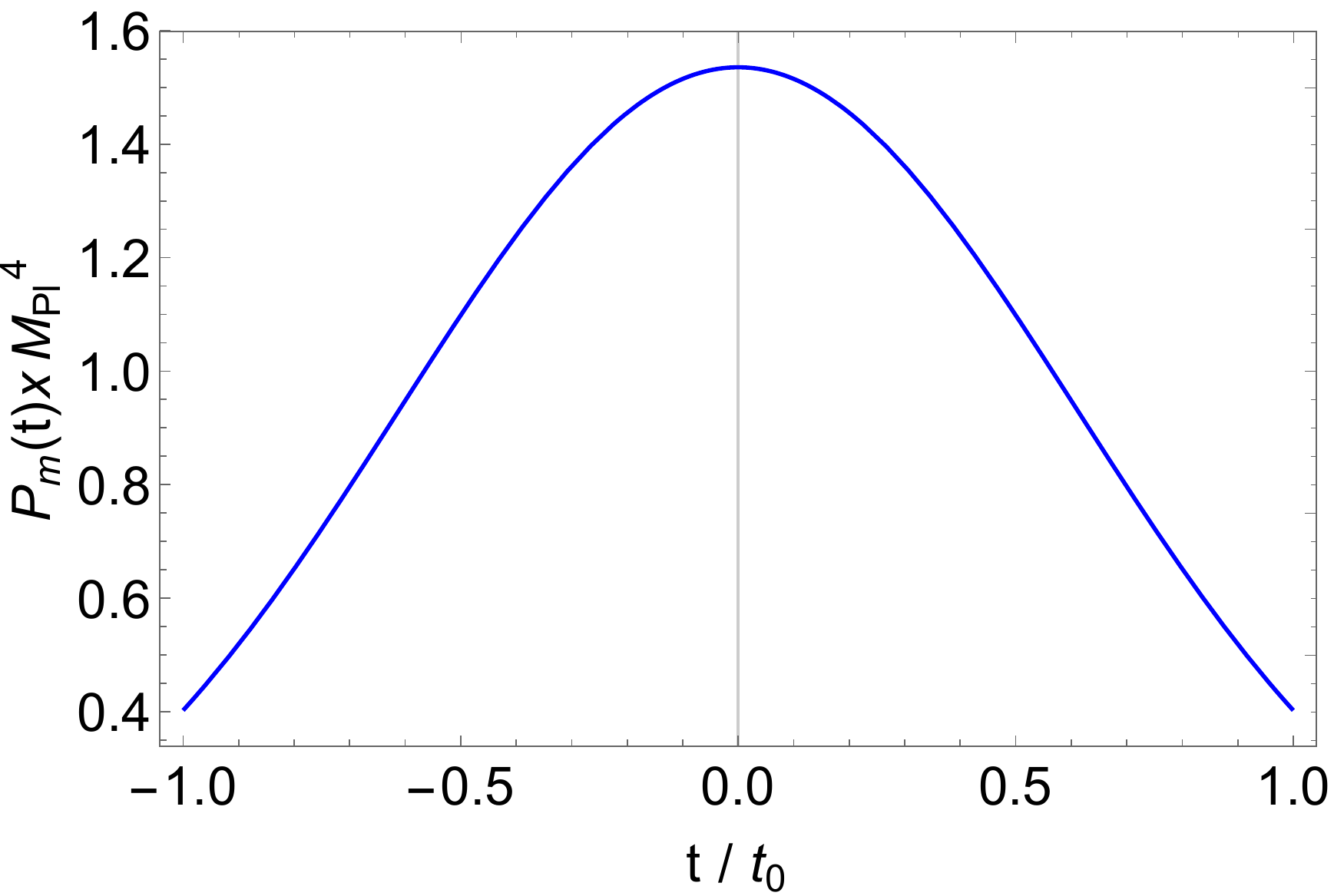}
         \caption{}
         \label{psb}
     \end{subfigure}
  \caption{(a) Time evolution of the scale factor $a(t)$, (b) the Hubble parameter $H(t)$, (c) Energy density $\rho_m$ of the KR field, (d) Matter pressure $p_m$ of the KR field in symmetric bounce for $\beta=7.46$ $\times 10^{-85}$ GeV$^2$ and $t_0= 6.7 \times 10^{41}$ GeV$^{-1}$}
     \label{fig:symmetricbounce}
     \end{figure*}

where $f_T = \frac{\partial f}{\partial T}$ and $f_{TT} = \frac{\partial^2 f}{\partial T^2}$.\\
Varying the KR action with respect to the field $B_{\mu \nu}$, gives the equation of motion as
\begin{equation}
    \stackrel{\circ}{\nabla}_{\mu}H^{\mu \nu \rho} = 0 \ .
    \label{KReqmotion}
    \end{equation}
The completely antisymmetric three-form field $H^{\mu \nu \rho}$ is physically equivalent to its Hodge dual, namely a one-form field in 4 dimensions. One can think of the one-form field to be following from a scalar potential $\phi$ and is defined as,
\begin{equation}
    H^{\mu \nu \lambda}=\varepsilon^{\mu \nu \lambda \rho}\partial_\rho \phi \ .
    \label{KRansatz}
\end{equation}
This however makes the above EOM second order in $\phi$ and can be written as
\begin{equation}
\begin{aligned}
     \stackrel{\circ}{\nabla}_{\mu}H^{\mu \nu \rho} 
     =\varepsilon^{\mu \nu \lambda \rho}\partial_\mu\partial_\rho \phi - \varepsilon^{\mu \nu \lambda \rho}\big({\Gamma}^{\sigma}{}_{\mu \rho}-K^{\sigma}{}_{\mu \rho}\big)\partial_\sigma \phi=0 \ .
  \end{aligned}
\end{equation}
The equation of motion of $\phi$ can now be obtained from the Bianchi identity
\begin{equation}
     \stackrel{\circ}{\nabla}_{[\mu}H_{ \nu \rho \sigma]} = 0  \ .
     \label{bianchi}
\end{equation}
Substituting Eq.\eqref{KRansatz} in Eq.\eqref{bianchi} and using the fact $\stackrel{\circ}{\nabla}_{[\mu}\varepsilon_{ \nu \rho \sigma \lambda]}\partial^\rho \phi = 0$ in four dimensions, we obtain the equation of motion of $\phi$ as
\begin{equation}
    \stackrel{\circ}{\nabla}_{\lambda}\partial^\lambda \phi=0 \ .
    \label{eomII}
\end{equation}
Using Eq.\eqref{KRansatz}, the equation of motion
\eqref{eom1} becomes
\begin{equation}
        M_\mu{}^\nu = h\kappa^2\Big(3\delta_\mu{}^{\nu}\partial^\rho \phi \partial_\rho \phi - 6\delta_\lambda{}^{\nu}\partial^\lambda \phi \partial_\mu \phi\Big)
        \label{eomI}
\end{equation}
Since we are interested in how the KR field affects the time evolution of the Universe, for simplicity, we consider $\phi$ as a function of the cosmic time $t$, satisfying the initial conditions
\begin{equation}
    \phi(t_b)=0 \ , \ \ \ \ \ \ \phi'(t_b)=1
    \label{initialcond}
\end{equation}
where $t_b$ is the time when the bounce occurs.
Now, the equations of motion Eq.\eqref{eomII} and Eq.\eqref{eomI} takes the form, 
\begin{align}
\label{eomfinal1}
 3H^2-6H^2f_T+\frac{1}{2}\big(f+\Lambda\big) &= \kappa^2\rho_m \ ,\\
 \label{eomfinal2}
 3H^2+2H'+\frac{1}{2}\big(f+\Lambda\big)-6H^2f_T-&2H'f_T-2Hf_{TT}T'\nonumber \\ = -\kappa^2p_m \ ,\\
 \label{eomphi}
\phi''+3H\phi'&=0 \ ,
 \end{align}
where $\rho_m$ and $p_m$ are the energy density and the matter pressure of the Kalb-Ramond field in the Universe, given by
\begin{equation}
     \rho_m = 3\phi'^2\ , \ \ \ \ \ \ 
    p_m = 3\phi'^2a^2 \ .
    \label{density}
\end{equation}
Equations \eqref{eomfinal1} and \eqref{eomfinal2} can be together written as
\begin{equation}
    2H'-2H'f_T-2Hf_{TT}T' = -3\kappa^2\phi'^2(a^2+1) \ .\\
    \label{eqfinal}
\end{equation}
The  Eq.\eqref{eomphi} then gives the solution of the KR field as
\begin{equation}
    \phi(t)=\int_1^t e^{\big(-\int_1^\zeta 3H(\xi)d\xi\big)}c_1d\zeta + c_2 \ ,
\end{equation}
where $c_1$ and $c_2$ are constants set to satisfy the initial conditions Eq.\eqref{initialcond}.  \\
\\
%

In particular, we will be looking into two cases of nonsingular bouncing cosmology, namely
\begin{enumerate}
    \item[A.] Symmetric bounce
    \item[B.] Matter bounce
\end{enumerate}
\subsection{Symmetric bounce}
\begin{figure}
    \centering
    \includegraphics[scale=0.45]{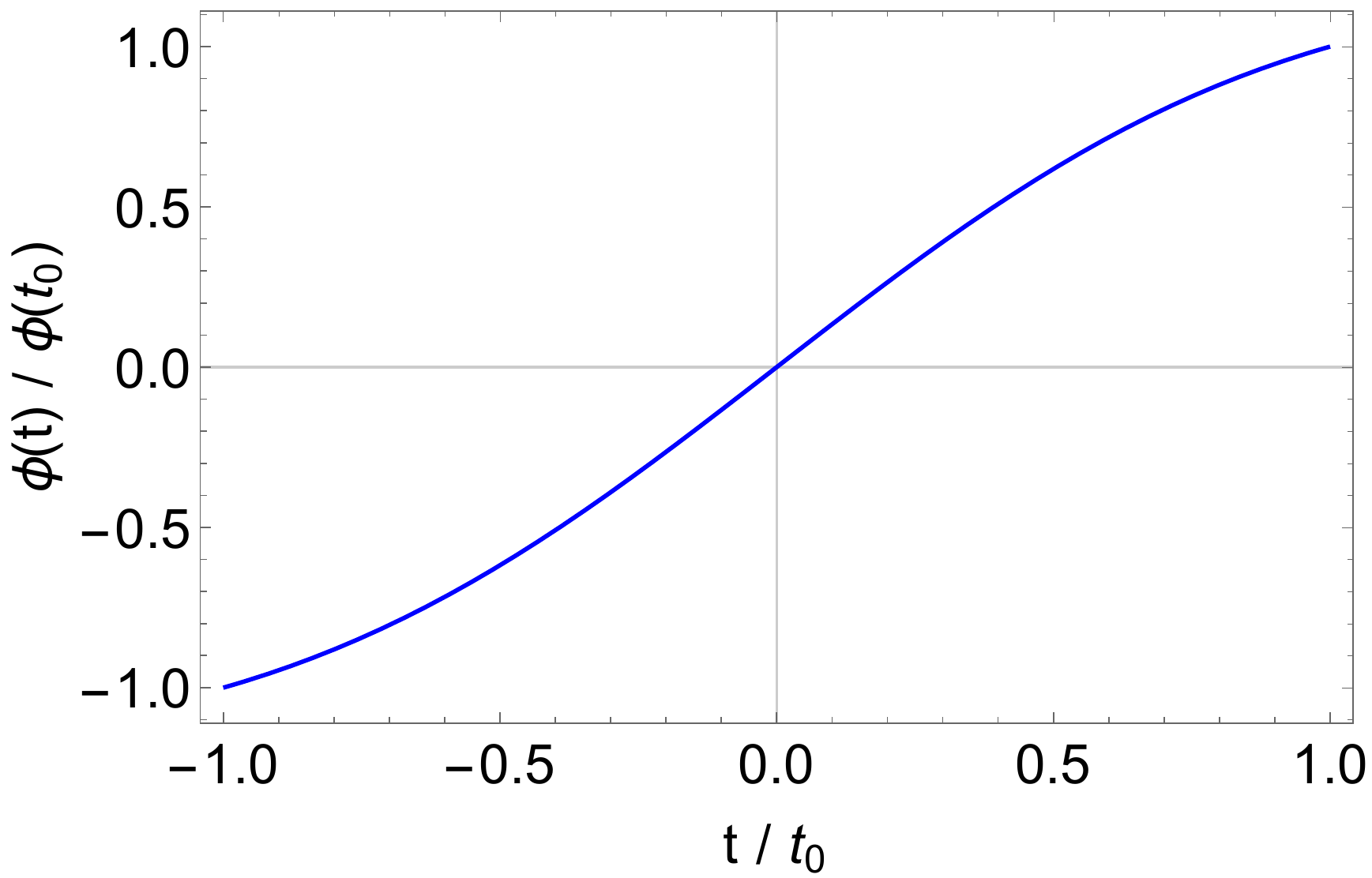}
    \caption{Time evolution of the scalar field $\phi(t)$ in symmetric bounce for $\beta=7.46$ $\times 10^{-85}$ GeV$^2$ and $t_0= 6.7 \times 10^{41}$ GeV$^{-1}$.}
    \label{phi(t)}
\end{figure}

In symmetric bouncing cosmology, the scale factor is given as \cite{Cai:2012va, Caruana:2020szx}
\begin{equation}
    a(t)=a_0\operatorname{exp}\Big(\alpha \frac{t^2}{t_*^2}\Big) \ ,
    \label{scale} 
\end{equation}
where $a_0 = a(0) > 0$ is the minimum value attained by the scale factor, $t_* > 0$ is an arbitrary time and $\alpha>0$ is a parameter. Fig. \eqref{a(t)sb} shows the behavior of $a(t)$ over time, where we chose the parameter $\beta$ = $\alpha/t_*^2$. There is a particular time $t_0>0$ when the scale factor becomes unity  i.e $a(t_0)=1$. We define $t_0$ to be the present cosmological time with the present Hubble parameter $H_0 \equiv H(t_0)$. The expression for $t_0$ using Eq.\eqref{scale} is given as
\begin{equation}
    t_0 = \sqrt{\frac{-\operatorname{ln}a_0}{\beta}} \ .
    \label{t0}
\end{equation}
 Since $\beta >0 $,  Eq.\eqref{t0} restricts the range $a_0$ $\in$ $(0,1)$. The current time is computed to be $t_0= 6.7 \times 10^{41}$ GeV$^{-1}$, according to the Planck Collaboration results 2015 \cite{Planck:2015fie}. Given the expression of the scale factor, it is straightforward to calculate the Hubble parameter and the torsion scalar as
\begin{equation}
    H(t) = 2\beta t \ , \ \ \ \ \ \  T(t) =24\beta^2 t^2 \ .
    \label{TH}
\end{equation}
In Fig.\eqref{h(t)sb}, we plot the Hubble parameter over time, where $H(t)$ varies linearly with time. The Hubble parameter's positivity determines whether a Universe is expanding or contracting. The phase when $H<0$ for $t<0$ is the contracting phase followed by the expansion phase where $H>0$ for $t>0$.
Clearly, the bounce occur at $t=0$ (which is a nonsingular bounce), when $H=0$. The current observational value of Hubble constant is $H_0$ = $H(t_0) \approx 10^{-42}$ GeV. Using $H_0$ and $t_0$ in Eq. \eqref{TH}, we get the value of $\beta$ to be $7.46$ $\times 10^{-85}$ GeV$^2$. 
Solving the equation of motion  \eqref{eomphi} using the initial conditions Eq.\eqref{initialcond}, we get the expression of $\phi$ in the symmetric bounce cosmology,
\begin{equation}
    \phi(t)= \frac{1}{2}\sqrt{\frac{\pi}{3\beta}}\operatorname{erf}\Big(\sqrt{3\beta}t\Big) \ ,
    \label{phidot}
\end{equation}
where $\operatorname{erf}(x)$ is the error function. In Fig.\eqref{phi(t)}, we plotted the time evolution of $\phi(t)$. $\phi(t)$ behaves as a sigmoid function, varying monotonically, but almost saturates after a certain point. This is evident from the asymptotic behavior of $\phi(t)$,
\begin{equation}
    \operatorname\lim_{t\to \infty}\phi(t)=\frac{1}{2}\sqrt{\frac{\pi}{3 \beta}} \ .
\end{equation}
The energy density and pressure of the KR field can be obtained using Eq.\eqref{density} as
\begin{align}
    \rho_m &= 3 \operatorname{exp}\Big(-6\beta t^2\Big) \ ,\\
    p_m & = 3\operatorname{exp}\big(-2\beta t_0^2\big) \operatorname{exp}\Big(-4\beta t^2\Big)\ .
    \label{edsb}
\end{align}

\begin{figure*}
     \centering
     \begin{subfigure}[b]{0.44\textwidth}
         \centering
         \includegraphics[scale=0.4]{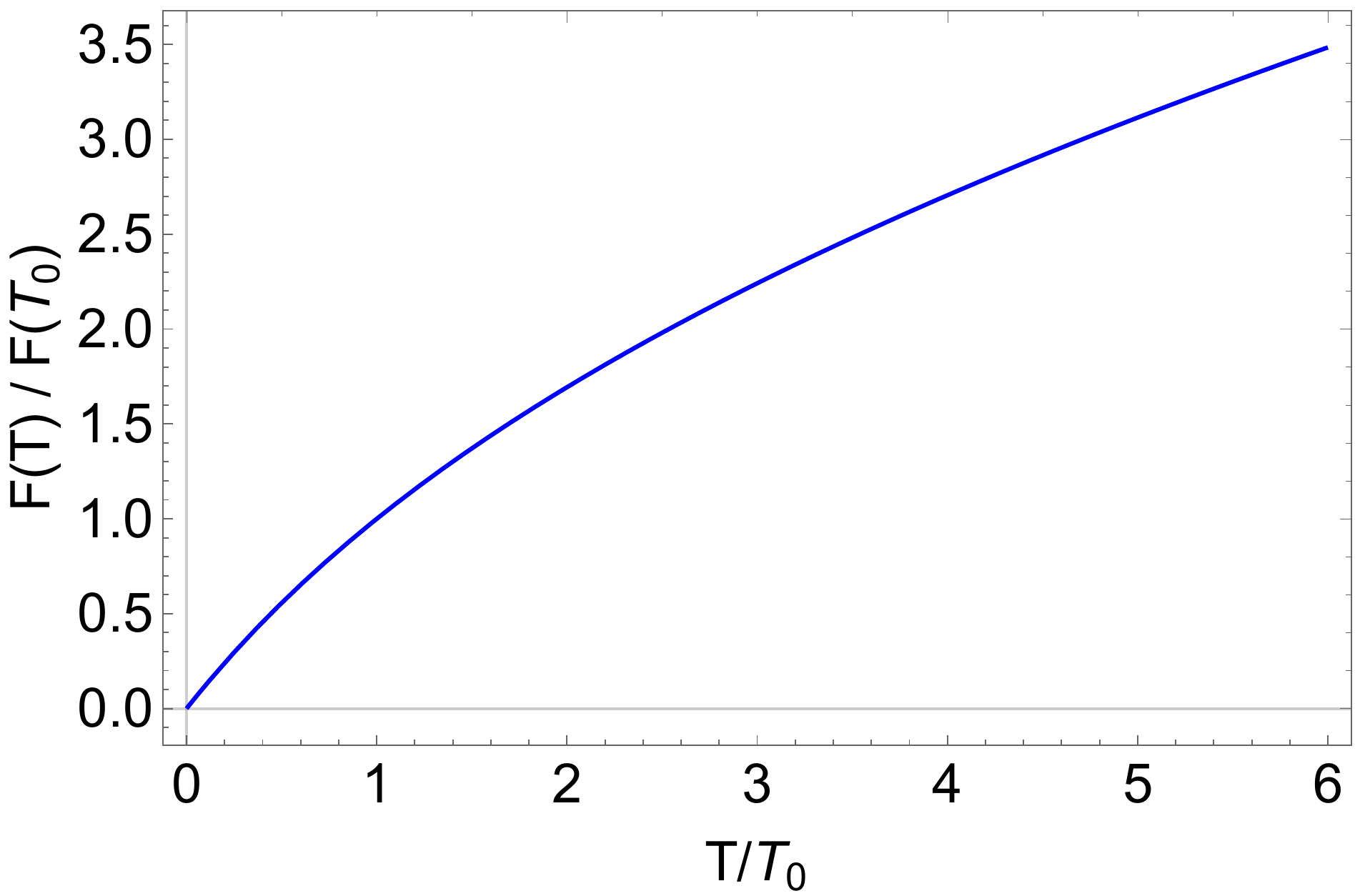}
         \caption{}
         \label{fTsb}
     \end{subfigure}
     \hfill
     \begin{subfigure}[b]{0.4\textwidth}
         \centering
         \includegraphics[scale=0.4]{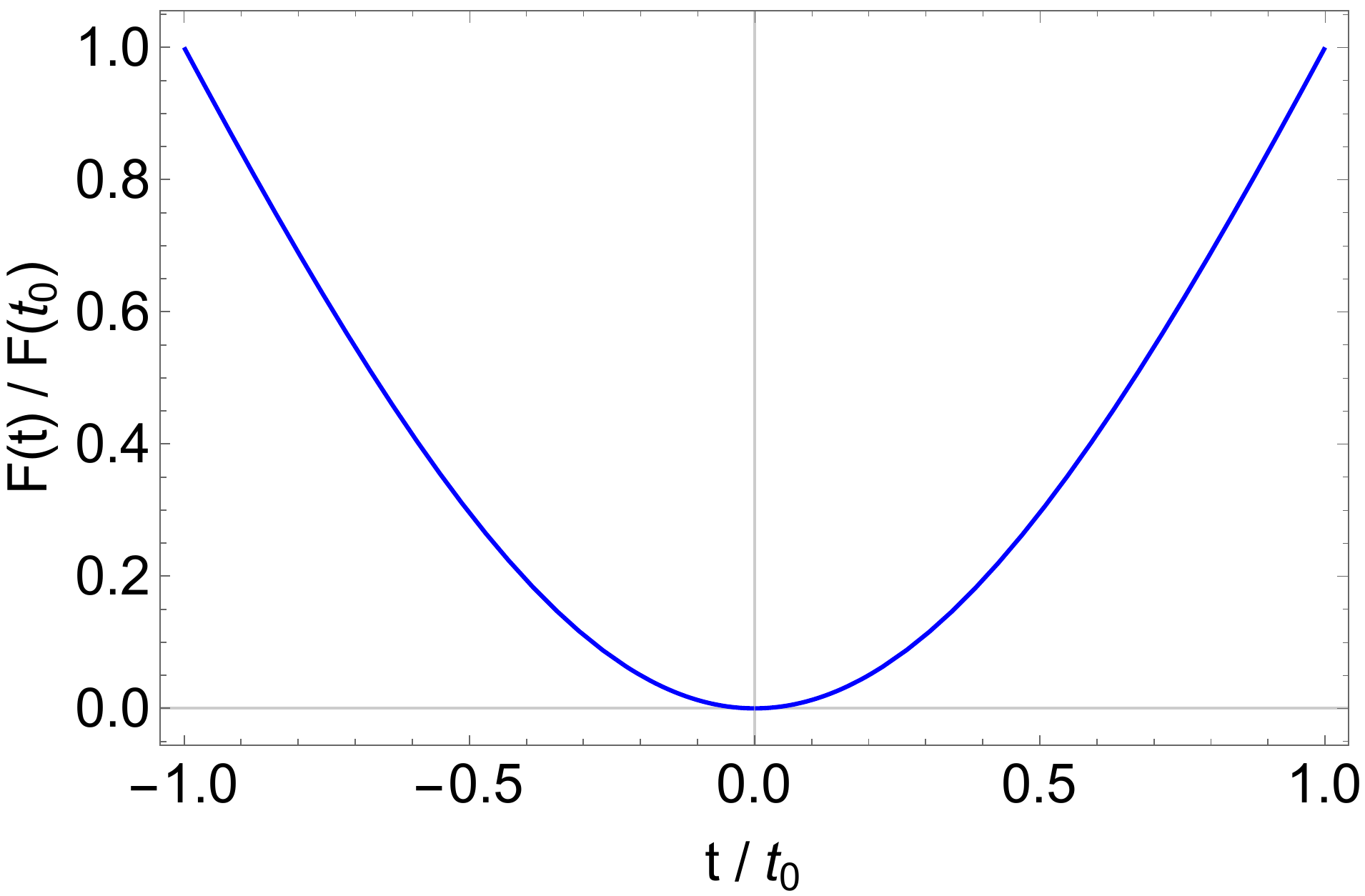}
         \caption{}
         \label{ftsb}
     \end{subfigure}
     \caption{The plot $(a)$ shows $F(T)$ vs $T/T_0$ in symmetric bounce scenario for $\beta=7.46 \times 10^{-85}$ GeV$^2$, which corresponds to the present Universe. Here, $T_0=24\beta \operatorname{ln}a_0$ is the torsion scalar at $t=0$ and we have chosen  $\kappa=M_{Pl}$. The evolution of $F(T)$ with respect to  $t/t_0$ is plotted in $(b)$}
     \label{F(T)}
     \end{figure*}

The evolution of energy density and matter pressure with respect to the cosmic time $t$ is plotted in Fig.\eqref{rhosb} and Fig.\eqref{psb} respectively. Both the plots show a similar behavior with bell-like profile and localization at $t=0$. Further,it is evident that the evolution depends on the factor $\beta$, which determines how fast the Universe expands or contracts. The energy density at the bounce is obtained to be $3$ M$_{Pl}{}^4$, and at the present time $t_0$, it is  $1.34$ M$_{Pl}{}^4$. Similarly the matter pressure $p_m$ at $t=0$ and $t=t_0$ are $1.53$ M$_{Pl}{}^4$ and $0.9$ M$_{Pl}{}^4$ respectively. Clearly, the localization of energy densities at $t=0$ is responsible for the bounce, but is large enough to have its effects noticeable at the present day cosmology. 
Using Eq.\eqref{eqfinal}, we get the differential equation of f(T) as
\begin{equation}
\begin{aligned}
    2Tf_{TT}+f_T = & 1 + \frac{3}{4 \beta}\operatorname{exp}\big(-\beta t_0^2\big)\kappa^2\operatorname{exp}\Big(\frac{-T}{6\beta}\Big)\\ &+ \frac{3}{4 \beta}\kappa^2\operatorname{exp}\Big(\frac{-T}{4 \beta}\Big) \ ,
    \end{aligned}
\end{equation}
Solving the above differential equation, we finally derive the exact functional form of $F(T)$ to be
\begin{equation}
\begin{aligned}
    F(T) & = \frac{3}{2}\kappa^2\operatorname{exp}\Big(\frac{-T}{4 \beta}\Big)\Big[2+3\operatorname{exp}\big(-\beta t_0^2\big)\operatorname{exp}\Big(\frac{T}{12 \beta}\Big)\Big]\\ & +6\sqrt{\pi}\sqrt{\frac{T}{\beta}}\kappa^2\operatorname{erf}\Big(\frac{1}{2}\sqrt{\frac{T}{\beta}}\Big)\\ &+3\sqrt{6\pi}\operatorname{exp}\big(-\beta t_0^2\big)\sqrt{\frac{T}{\beta}}\kappa^2\operatorname{erf}\Big(\sqrt{\frac{T}{6 \beta}}\Big) + C \ .
    \label{FTsymm}
    \end{aligned}
\end{equation}
where $C$ is an integration constant. 
Moreover it is important to note that the reconstructed Lagrangian Eq.\eqref{FTsymm} is an even function, and hence is symmetric with respect to the bounce at $t=0$. \\
\\
\begin{figure*}
     \centering
     \begin{subfigure}[b]{0.44\textwidth}
         \centering
         \includegraphics[scale=0.45]{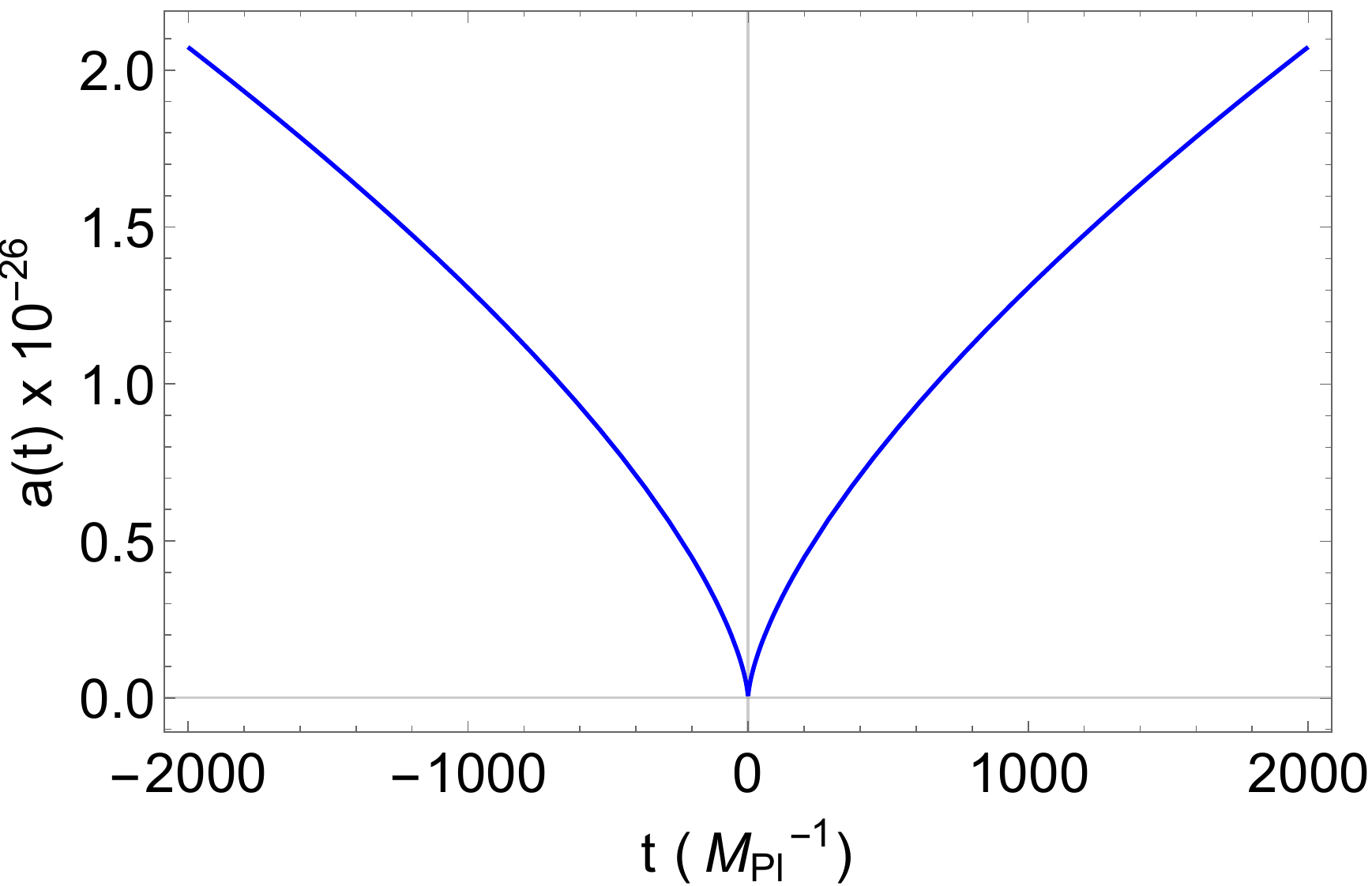}
         \caption{}
         \label{a(t)mb}
     \end{subfigure}
     \hfill
     \begin{subfigure}[b]{0.44\textwidth}
         \centering
         \includegraphics[scale=0.45]{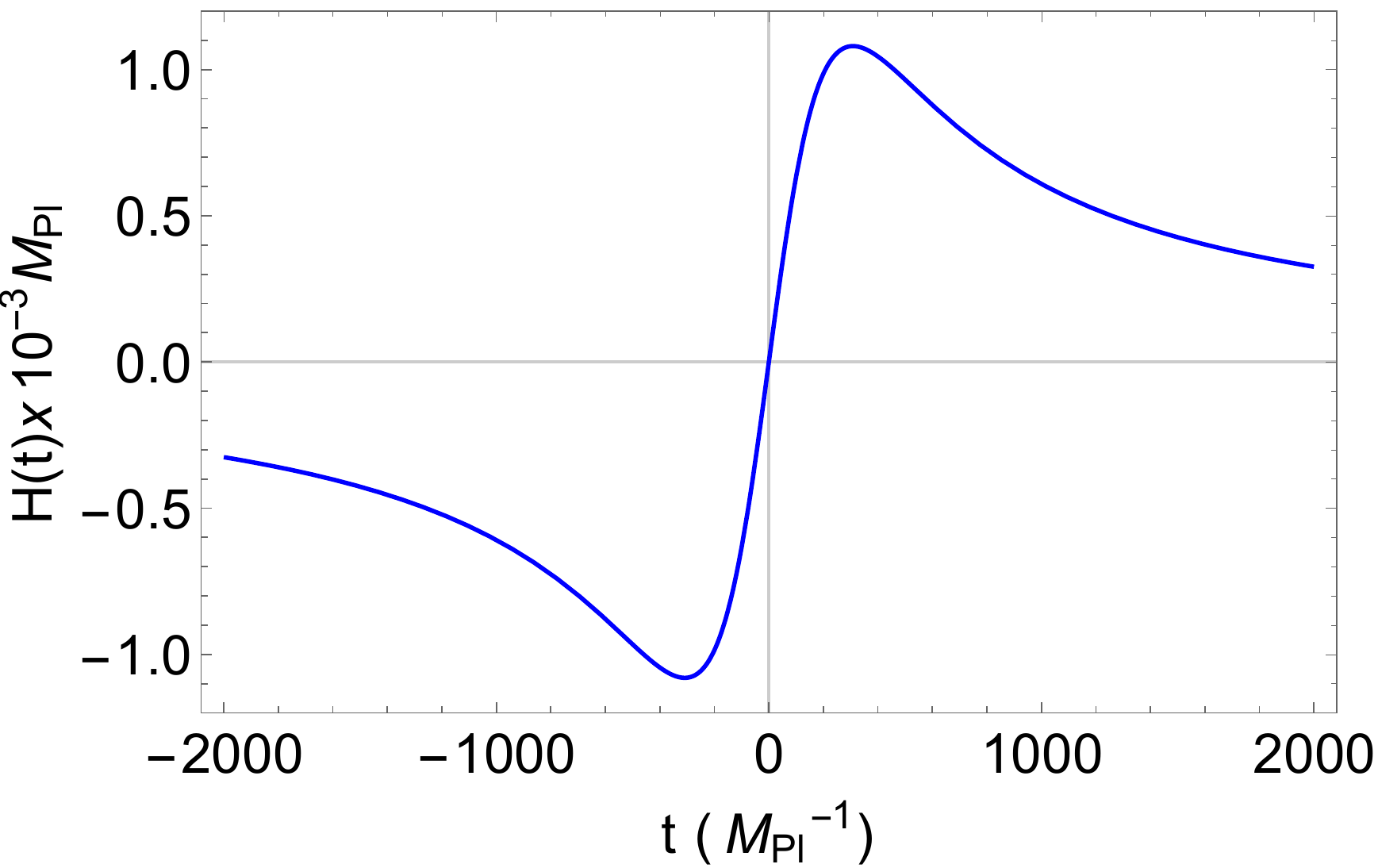}
         \caption{}
         \label{h(t)mb}
     \end{subfigure}
     \newline
     \begin{subfigure}[b]{0.44\textwidth}
         \centering
         \includegraphics[scale=0.45]{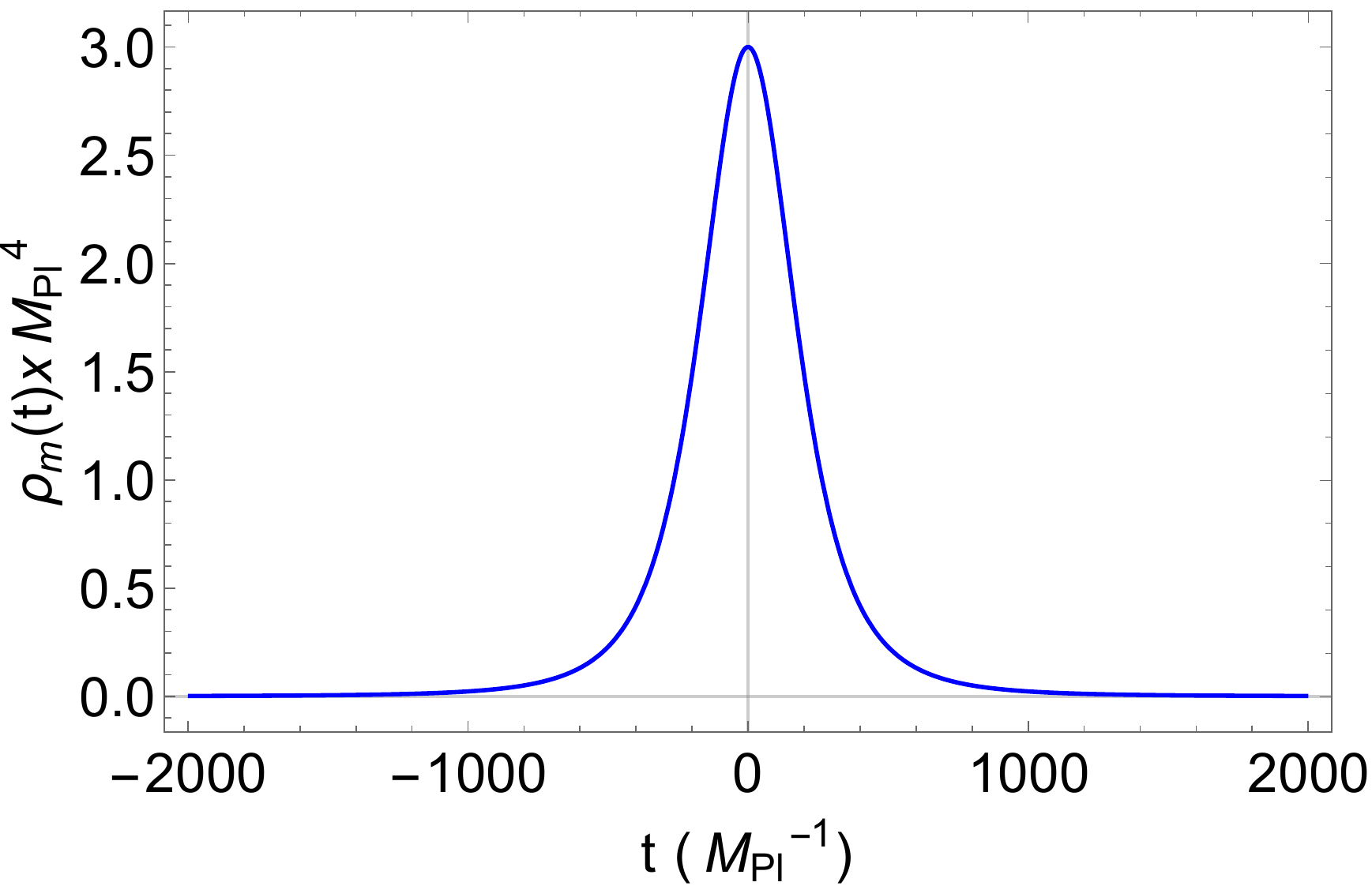}
         \caption{}
         \label{r(t)mb}
     \end{subfigure}
     \hfill
     \begin{subfigure}[b]{0.44\textwidth}
         \centering
         \includegraphics[scale=0.45]{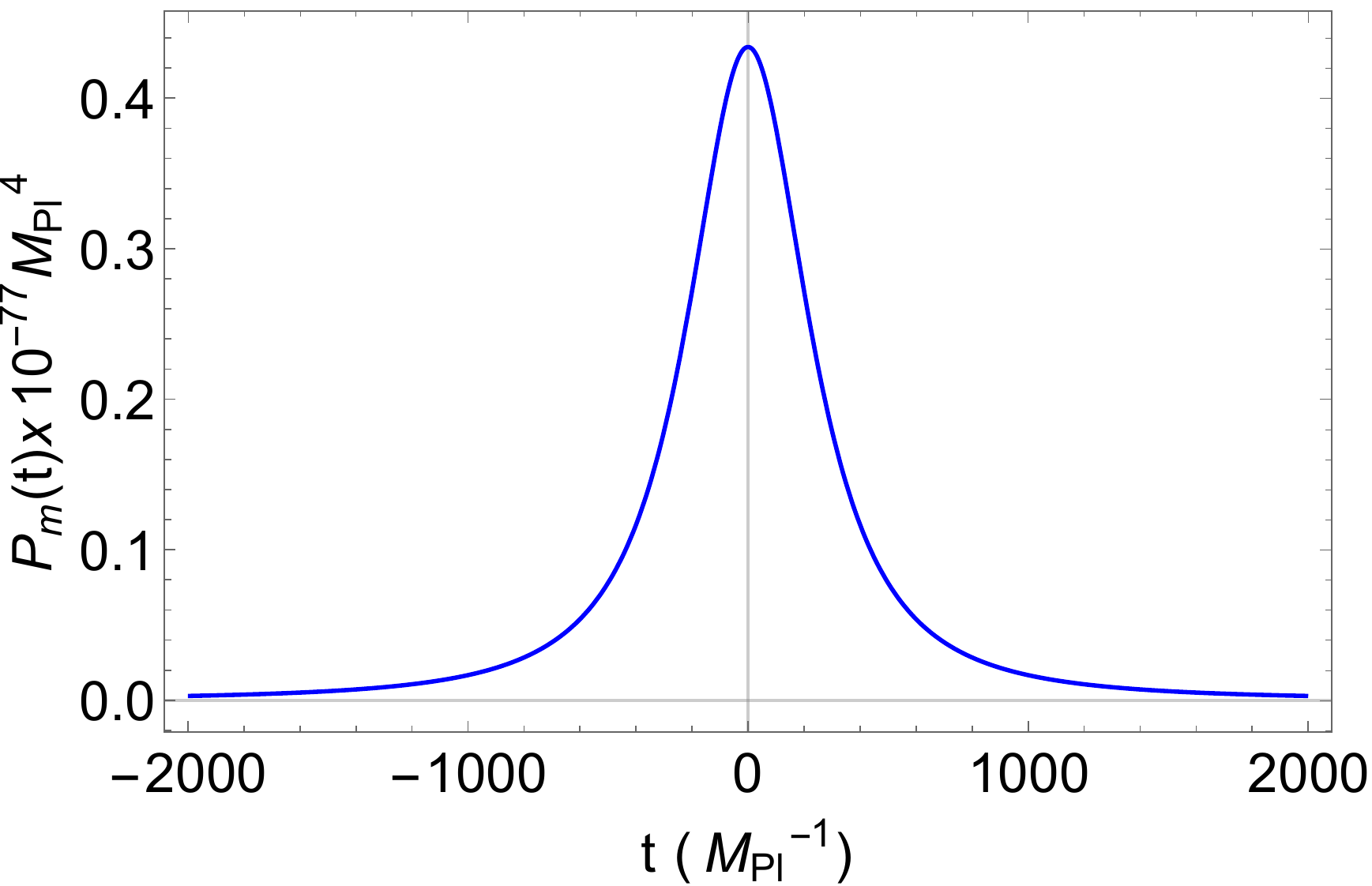}
         \caption{}
         \label{p(t)mb}
     \end{subfigure}
  \caption{(a) Time evolution of the scale factor $a(t)$, (b) the Hubble parameter $H(t)$, (c) Energy density of the KR field, (d) Matter pressure of the KR field in matter bounce for $\sigma=7\times 10^{-6}M^2_{Pl}$ and $t_0= 6.7 \times 10^{41}$ GeV$^{-1}$}
\label{fig:matterbounce}
\end{figure*}
For the gravitational Lagrangian to be able to recover vacuum solutions, $T$ has to be zero in the absence of matter \cite{Ferraro:2011ks}. This is evident from  Eq.\eqref{TH}. Also, as a consequence of Eq.\eqref{eomfinal1} we assume $\Lambda=6\kappa^2$ such that it satisfies the  vacuum solution constraint $f(0)=0$. This fixes the integration constant C to be
\begin{equation}
    C=-\Big(3+\frac{9}{2 \operatorname{exp}\big(\beta t_0^2\big)}\Big) \ .
\end{equation}
Figure.\eqref{fTsb} shows the function $F(T)$ vs torsion scalar $T$ and Fig.\eqref{fTsb} shows the evolution of $F(T)$ with respect to the cosmic time $t$, corresponding to the symmetric-bounce scenario in the presence of Kalb-Ramond field described by Eq.\eqref{phidot}. 
\subsection{Matter bounce}

\begin{figure}[]
    \centering
    \includegraphics[scale=0.45]{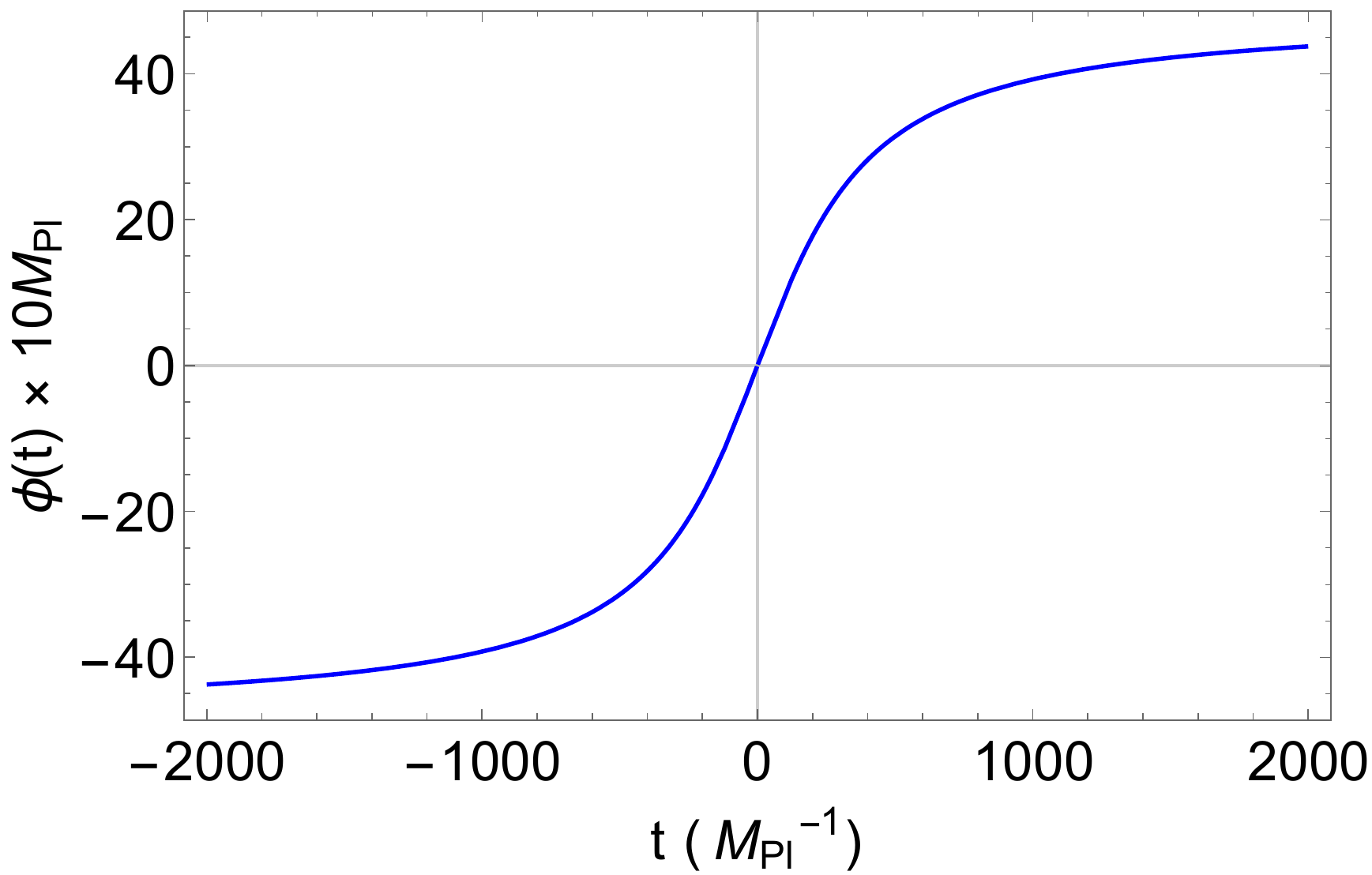}
    \caption{Time evolution of the scalar field $\phi(t)$ in matter bounce for $\sigma=  7\times 10^{-6}M^2_{Pl}$ and $t_0= 6.7 \times 10^{41}$ GeV$^{-1}$.} 
    \label{phi(t)mb}
\end{figure}

In matter bounce cosmology \cite{Singh:2006im, Wilson-Ewing:2012lmx, Caruana:2020szx} the scale factor is given as
\begin{equation}
    a(t)= a_0 \Big(\frac{3}{2}\sigma t^2 +1 \Big)^\frac{1}{3} \ ,
    \label{a(t)matter}
\end{equation}
where $a(0)$= $a_0$ is a positive quantity, and $0<\sigma << 1$ is the a positive quantity, which is determined from the loop quantum gravity \cite{Caruana:2020szx}. The parameter $\sigma$ also determines how fast the bounce occurs \cite{Cai:2011tc}. Fig.\eqref{a(t)mb} shows the time evolution of the scale factor in matter bounce cosmology.
The present cosmological time $t_0>0$ can be obtained from Eq.\eqref{a(t)matter} as
\begin{equation}
   t_0 = \sqrt{\frac{2}{3\sigma}\Big(\frac{1}{a_0^3}-1\Big)} \ .
   \label{presentctimemb}
\end{equation}
Thus, the range of $a_0$ is restricted to $(0,1)$, since $\sigma>0$. We have taken $\sigma=7\times 10^{-6}M^2_{Pl}$, which is determined by the amplitude of the CMB spectrum \cite{Cai:2011tc} and the present Hubble constant could be evaluated using this $\sigma$ to be $H_0 \approx 10^{-42}$ GeV.
The expressions of the Hubble parameter and the torsion scalar in matter bounce cosmology takes the form,
\begin{equation}
    H(t) = \frac{2\sigma t}{3\sigma t^2+2} \ \ \ \ \ \  T(t) = \frac{24\sigma^2t^2}{(3\sigma t^2+2)^2} \ .
    \label{THmb}
\end{equation}
$H(t)$ is plotted in Fig \eqref{h(t)mb}, which clearly shows that a nonsingular bounce occurs at $t=0$, with a contracting and expansion phase for $t<0$ and $t>0$ respectively. The torsion scalar at the cosmic time $t_0$ can be obtained by substituting Eq.\eqref{presentctimemb} in Eq.\eqref{THmb}, which is given as
\begin{equation}
 T_0 \equiv T(t_0) = 4 a_0^3 \left(1-a_0^3\right) \sigma \ .
 \label{T0mb}
\end{equation}
The corresponding energy density and matter pressure of the KR field is obtained as
\begin{align}
    \rho =&  \frac{12}{(3 \sigma t^2+2)^2} \ ,\\
    p =& \frac{6 a_0^2}{2^{-\frac{1}{3}}\big(3 \sigma t^2+2\big)^\frac{4}{3}} \ .
\end{align}
These are plotted in Fig.\eqref{r(t)mb} and Fig. \eqref{p(t)mb}, which shows that the maximum of energy density and matter pressure is again at $t=0$. At the bounce, the energy density is $3$ M$_{Pl}{}^4$, but it drastically decreases to $6.1 \times 10^{-234}$ M$_{Pl}{}^4$ at the present time $t=t_0$. This feature explains the lack of cosmological effects of the KR field in the present day Universe. Upon solving the KR field equation [Eq.\eqref{eqfinal}] we get the expression of the scalar field $\phi$ corresponding to the matter bounce cosmology as
\begin{equation}
    \phi(t)= \sqrt{\frac{2}{3\sigma}}\operatorname{tan^{-1}}\Big(\sqrt{\frac{3\sigma}{2}}t\Big) \ .
    \label{phi(t)matter}
\end{equation}
The time evolution of $\phi(t)$ is plotted in the Fig.\eqref{phi(t)mb}. It can be observed that the behaviour of $\phi$ is again similar to what we have seen in the case of symmetric bounce, where it behaves as a sigmoid function. The asymptotic behavior of $\phi(t)$ in matter bounce as $t\rightarrow \infty$ is given as
\begin{equation}
    \operatorname\lim_{t\to \infty}\phi(t)=\frac{\pi}{\sqrt{6\sigma}} \ .
\end{equation}
Solving the functional form of $F(T)$ using  Eq.\eqref{phi(t)matter} in Eq.\eqref{eomfinal1} and Eq.\eqref{eomfinal2},  we get $F(T)$ as a function of $t$,
\begin{equation}
    \begin{aligned}
       F(t) &= \frac{24\kappa^2}{(2+3t^2\sigma)^2}-\frac{48t^2\sigma^2}{(2+3t^2\sigma)^2}\\ + &  \frac{6t^2\sigma}{(2+3t^2\sigma)^2}\Bigg[8\sigma+\kappa^2\Big(6+9a_0^2\big(\sqrt{2}(2+3t^2\sigma)\big)^\frac{2}{3}\Big)\Bigg]\\+&\frac{6\sqrt{6}t\sigma^\frac{1}{2}}{(2+3t^2\sigma^2)}\Bigg[\kappa^2\operatorname{tan^{-1}}\Big(\sqrt{\frac{3\sigma}{2}}t\Big)\Bigg]\\ - & \frac{18a_0^2\kappa^2t^2\sigma}{(2+3t^2\sigma^2)}\operatorname{{}_2F_1}\Big[\frac{1}{3},\frac{1}{2};\frac{3}{2};-\frac{3t^2\sigma}{2}\Big] -6\kappa^2 \ ,
    \end{aligned}
    \label{F(t)mbequation}
\end{equation}
where ${{}_2F_1}[a,b;c;d]$ represents the hypergeometric function. \\

\begin{figure*}
     \centering
     \begin{subfigure}[b]{0.45\textwidth}
         \centering
         \includegraphics[scale=0.45]{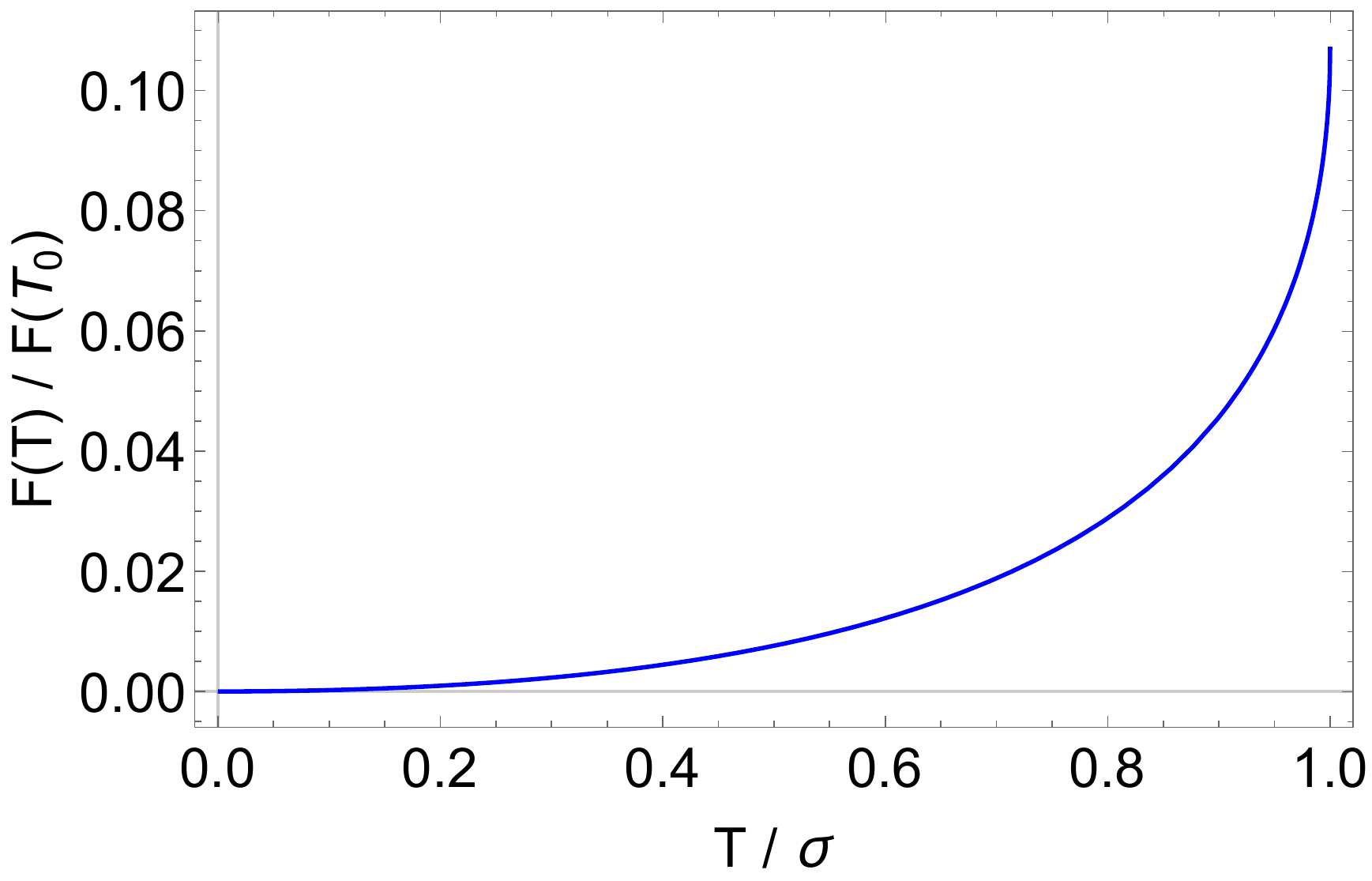}
         \caption{}
         \label{f(T)mb}
     \end{subfigure}
     \hfill
     \begin{subfigure}[b]{0.44\textwidth}
         \centering
         \includegraphics[scale=0.44]{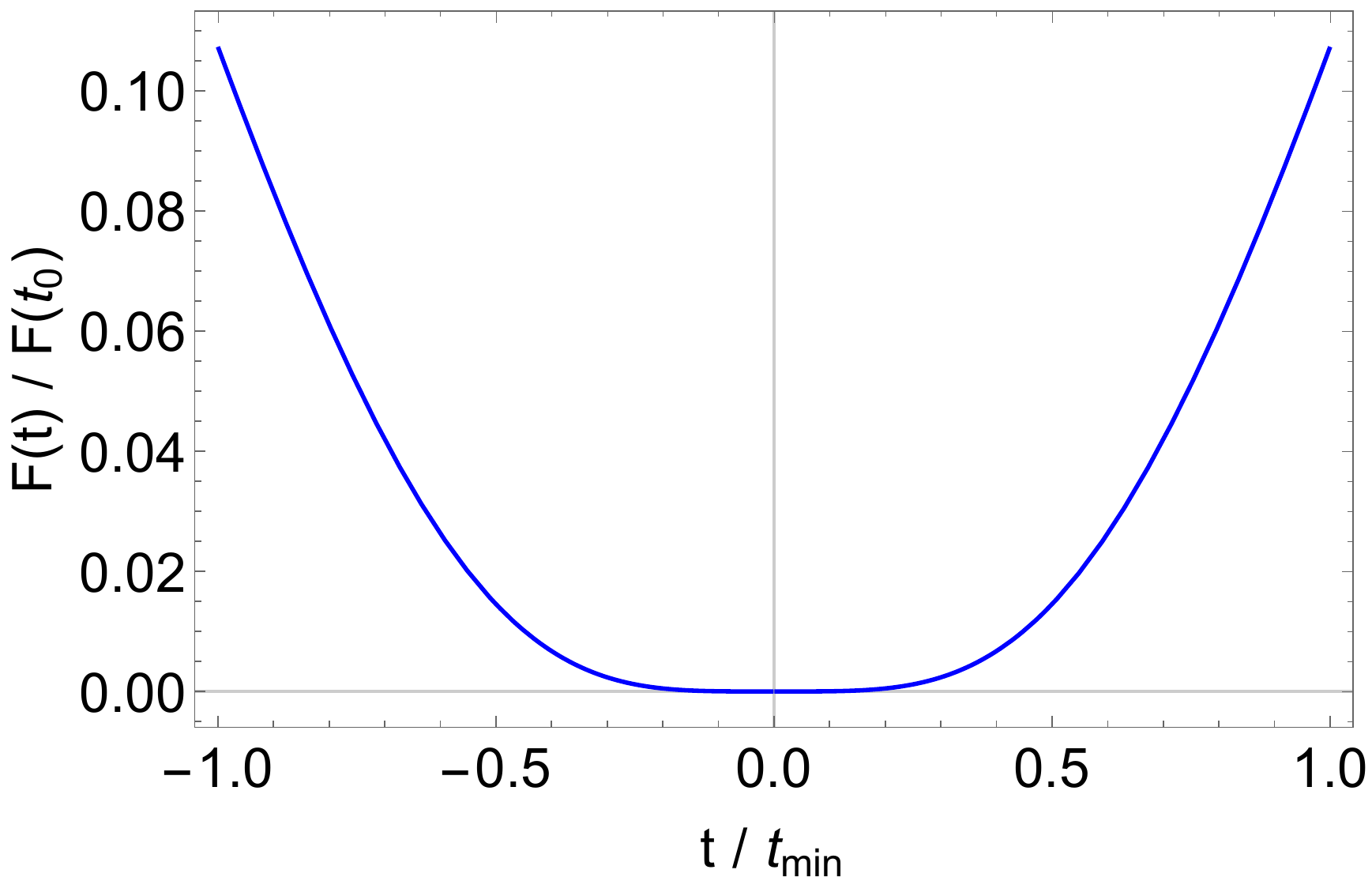}
         \caption{}
         \label{f(t)mb}
     \end{subfigure}
     \caption{The plot $(a)$ shows function $F(T)$ in terms of $T/\sigma$ in the matter bounce scenario In (a), $F(T)$ is only valid for $T/\sigma\leq1$, equivalently $|t| \leq t_{min}$=$\sqrt{\frac{2}{3\sigma}}$ in (b). In plot (b), evolution $F(T)$ in terms of the  $t/t_{min}$ for $\sigma=7\times 10^{-6}M^2_{Pl}$ is plotted}
     \label{F(T)mb}
     \end{figure*}

The symmetry between the contraction and the expansion phase in matter bounce requires $F(t)$ to be an even function of $t$. In Fig.\eqref{f(t)mb}, we have plotted the cosmic time evolution of $F(t)$, which shows its symmetric behavior with respect to the bouncing point at $t=0$. The inverse relation $t(T)$ can be obtained by the inversion of $T$ in Eq.\eqref{THmb},
 \begin{equation}
     t(T)= \pm \sqrt{\frac{2}{3}}\sqrt{\frac{2}{T}-\frac{1}{\sigma}-\frac{2}{T}\sqrt{1-\frac{T}{\sigma}}} \ .
     \label{tmb}
 \end{equation}
     
    Here, we have retained the solution pair that produces the desired result, $T=0$ at $t=0$. It is important to note that the solution is invertible only for $-\sqrt{\frac{2}{3\sigma}}\leq t \leq\sqrt{\frac{2}{3\sigma}}$. This corresponds to a characteristic time period for each matter bounce Universe corresponding to the critical parameter $\sigma$. Beyond this region of time, we assume TEGR is valid \cite{Cai:2011tc}. As the solution Eq.\eqref{F(t)mbequation} is an even function of t, both $\pm$ solutions in Eq.\eqref{tmb} provides the identical form of $F(T)$, with $-$ and $+$ solutions, representing the contraction and expansion phases respectively.
  
 Substituting Eq.\eqref{tmb} in Eq.\eqref{F(t)mbequation}, we get the functional form of F(T) as
\begin{equation}
\begin{aligned}
   & F(T)=\\ & \frac{1}{3h(T)^2\sigma^2} \Bigg[ 3 \kappa ^2 \Big(a_0^2 (T-2 \sigma  h(T)) \Big(3\times 2^{2/3} T \Big(\frac{\sigma  h(T)}{T}\Big)^{2/3}\\ &- 2 \sigma  h(T)  _2F_1\Big(\frac{1}{3},\frac{1}{2};\frac{3}{2};1-\frac{2 h(T) \sigma
   }{T}\Big)\Big)\\- &2 \sigma  T h(T) \Big(\sqrt{\sigma } \sqrt{\frac{2 h(T)}{T}-\frac{1}{\sigma }} \tan ^{-1}\Big(\sqrt{\sigma } \sqrt{\frac{2 h(T)}{T}-\frac{1}{\sigma }}\Big)\\ &+1\Big)\Big)\Bigg] - 6\kappa^2 \ \ \
\end{aligned}
\label{ft-matter}
\end{equation}
 which is also restricted to $T\leq\sigma$ (equivalently, $-\frac{2}{3\sigma}\leq t \leq\frac{2}{3\sigma}$).
In Eq.\eqref{ft-matter}, we have taken $h(T)$ as a functions of $T$, given as
 \begin{gather}
     h(T) = 1-\sqrt{1-\frac{T}{\sigma}} \ .
     \label{h(t)}
 \end{gather}
In Fig \eqref{f(T)mb}, we have plotted the function $F(T)$ in terms of $T$. Note that the solution Eq.\eqref{ft-matter} satisfies the vacuum solution constraint $F(0)=0$. 

\section{Discussions and conclusion}
\label{sec:discussion}
 
TEGR is a successful gravity description, specifically in the presence of sources that could twist the geometry to create torsion. One such example that could source torsion is the antisymmetric rank-2 Kalb-Ramond field. These antisymmetric tensor fields form an integral part of heterotic string models \cite{Kalb:1974yc,Cremmer:1973mg} as massless closed string modes and of some supersymmetric models like N=2 and N=8 extended SUGRA. They have also been widely studied \cite{Majumdar:1999jd} in the context of electromagnetic field coupling to the Einstein-Cartan system. Though essential, it is noteworthy that there are no experimental evidence for this field in the present day Universe \cite{Das:2018jey}. 

Since the presence of torsion breaks the $U(1)$ invariance of the gauge theory, it is important to introduce a suitable coupling prescription in teleparallel gravity. To do that successfully, we first define an equivalent of the covariant derivative called the Fock-Ivanenko derivative operator (FIDO) in teleparallel geometry. In Sec. \eqref{sec:FID}, we generalize FIDO to operate on any $n$-form tensor field in ($d+1$)-dimensional space, using the equations Eq.\eqref{mcoupling}, Eq.\eqref{spinconnection1} and Eq.\eqref{Jab-Bab1} and in particular on KR field. For completeness, we show the equivalence of Fock-Ivanenko derivative of the KR field in teleparallel gravity to the Levi-Civita covariant derivative in Einstien's gravity in Appendix \eqref{apa}.

We then compute the equations of motion and show that the dynamics of tetrad fields in teleparallel geometry are governed by the KR field. To keep the discussion general, we start with a generic function $F(T)$. And we considered the effect of this setting in producing two bouncing cosmologies, namely symmetric bounce and matter bounce. The absence of initial singularity in cosmological evolution has been a significant advantage of bouncing cosmologies over the inflationary paradigm. In these scenarios, the big bang is replaced by a continuous phase of expanding and contracting.

Note that the scale factor for both these scenarios are sourced by the localized energy density of the KR field as shown in Fig.\eqref{rhosb} and Fig.\eqref{r(t)mb}. These plots indicate the nature of energy density with time `$t$'. At the present time $t_0=6.7\times 10^{41} $ GeV$^{-1}$, the symmetric-bounce scenario predicts the KR field energy density of $\rho_m=1.34 M_{Pl}^4$. Where as matter bounce predicts a much smaller energy density $\rho_m \sim 0$. Thus the lack of cosmological evidence of KR field in the present day Universe strongly advocate matter bounce scenario over symmetric bounce.  

In the symmetric bounce, the generalized teleparallel gravity is an increasing function of $T$, for $T \in (0,\infty)$ as shown in Fig.\eqref{fTsb}, but for large $T$, $F(T)$ behaves linearly. Whereas in matter bounce, $F(T)$ again exhibits similar behavior, albeit the evolution is valid up to some value $T<\sigma$, as shown in Fig.\eqref{F(T)mb}. The analytical results are given in Table.\eqref{tab:my-table}. In both these scenarios, the scalar field shows a wave profile [Fig.\eqref{phi(t)} and \eqref{phi(t)mb}] and energy profile [Fig.\eqref{rhosb} and \eqref{r(t)mb}] of a `kink' which  could be interesting to study further. It is also interesting to wonder how an axion/pseudoscalar field will behave in the teleparallel setting, given the possible parity violations.
 
\begin{acknowledgments}
 \begin{table*}[]
\centering
\begin{tabular}{|P{3cm}|P{3cm}|P{3cm}|P{7cm}|}
\hline 
Model & $a(t)$ & $\phi(t)$&F(T)   \\[1ex] \hline \hline
Symmetric bounce      &   $a_0\operatorname{exp}\Big(\alpha \frac{t^2}{t_*^2}\Big)$     & $\frac{1}{2}\sqrt{\frac{\pi}{3\beta}}\operatorname{erf}\Big(\sqrt{3\beta}t\Big)$ &   $\frac{3}{2}\kappa^2\operatorname{exp}\Big(\frac{-T}{4 \beta}\Big)\Big[2+3\operatorname{exp}\big(\beta t_0^2\big)\operatorname{exp}\Big(\frac{T}{12 \beta}\Big)\Big] +6\sqrt{\pi}\sqrt{\frac{T}{\beta}}\kappa^2\operatorname{erf}\Big(\frac{1}{2}\sqrt{\frac{T}{\beta}}\Big) +3\sqrt{6\pi}\operatorname{exp}\big(\beta t_0^2\big)\sqrt{\frac{T}{\beta}}\kappa^2\operatorname{erf}\Big(\sqrt{\frac{T}{6 \beta}}\Big) -\Big(3+\frac{9}{2 \operatorname{exp}\big(\beta t_0^2\big)}\Big)$                             \\[3.5ex] \hline
 Matter bounce     &  $a_0 \Big(\frac{3}{2}\sigma t^2 +1 \Big)^\frac{1}{3}$      & $\sqrt{\frac{2}{3\sigma}}\operatorname{tan^{-1}}\Big(\sqrt{\frac{3\sigma}{2}}t\Big)$  &   $\frac{1}{3h(T)^2\sigma^2} \Bigg[ 3 \kappa ^2 \Big(a_0^2 (T-2 \sigma  h(T)) \Big(3\times 2^{2/3} T \Big(\frac{\sigma  h(T)}{T}\Big)^{2/3}- 2 \sigma  h(T)  _2F_1\Big(\frac{1}{3},\frac{1}{2};\frac{3}{2};1-\frac{2 h(T) \sigma
   }{T}\Big)\Big)- 2 \sigma  T h(T) \Big(\sqrt{\sigma } \sqrt{\frac{2 h(T)}{T}-\frac{1}{\sigma }} \times \tan ^{-1}\Big(\sqrt{\sigma } \sqrt{\frac{2 h(T)}{T}-\frac{1}{\sigma }}\Big) +1\Big)\Big)\Bigg] - 6\kappa^2 $                              \\ [3.5ex]\hline
\end{tabular}
\caption{The KR-field $\phi(t)$ and the reconstructed Lagrangian $F(T)$ corresponding to different setups for bouncing cosmology. The function $h(T)$ used in the matter bounce Lagrangian is defined as Eq.\eqref{h(t)}}
\label{tab:my-table}

\end{table*} 
M.T.A. acknowledges financial support of DST through INSPIRE Faculty Grant No. [DST/INSPIRE/04/2019/002507]. 
\end{acknowledgments}

\appendix
\section{Coupling prescription using Fock–Ivanenko derivative operator in the Riemannian geometry}
\label{apa}
In the framework of Riemannian geometry, the mathematical equivalent of Fock-Ivanenko derivative of the KR field $B^{ab}$, using Eq.\eqref{mcoupling} and Eq.\eqref{Jab-Bab1}, is given by
\begin{equation}
\begin{aligned}
    \mathcal{D}_\mu B^{ab} &= \partial_\mu B^{ab} - \frac{i}{2}\Omega^{c d}{ }_{\mu}\big(i(\delta^a_c\eta_{dg}-\delta^a_d\eta_{cg})\big)B^{gb}\\
    &-\frac{i}{2}\Omega^{c d}{ }_{\mu}\big(i(\delta^b_c\eta_{dg}-\delta^b_d\eta_{cg})\big)B^{ag}\\
    &=\partial_\mu B^{ab}+\Omega^{a }{ }_{c \mu}B^{c d} +\Omega^{b }{ }_{c \mu}B^{a c} \ .
    \end{aligned}
    \label{FI-Bab}
\end{equation}
Now, using Eq.\eqref{WD-tetrad} and Eq.\eqref{rel1} in Eq.\eqref{spinconnection1}, we can write $\Omega^{a b}{ }_{\mu}$ as
\begin{equation}
\Omega^{a b}{ }_{\mu}=h_{\rho}^{a} \Tilde{\nabla}_{\mu} h^{b \rho} \ .
\label{spinGR}
\end{equation}
This can be equivalently written as
\begin{gather}
\partial_{\mu} h^{a}{ }_{\nu}+\Omega_{b \mu}^{a} h^{b}{ }_{\nu}-\Tilde{\Gamma}^{\rho}{}_{\nu \mu} h
_{\rho}^{a}=0 \ .
\label{FI-tetrad1}
\end{gather}
Using the definition of the Fock-Ivanenko derivative of the tetrads,
\begin{equation}
\mathcal{D}_{\mu} h_{\nu}^{a}=\partial_{\mu} h^{a}{ }_{\nu}+\Omega_{b \mu} h^{b}{ }_{\nu} \ ,
\label{Fock-der-tetrad}
\end{equation}
we can rewrite Eq.\eqref{FI-tetrad1} as
\begin{gather}
\mathcal{D}_{\mu} h^{a}{ }_{\nu}=\Tilde{\Gamma}^{\rho}{ }_{\nu \mu} h^{a}{ }_{\rho} \ .
\label{FI-tetrad}
\end{gather}
Substituting Eq.\eqref{spinGR} and Eq.\eqref{spacet-lorentz1} in Eq.\eqref{FI-Bab} and using Eq.\eqref{FI-tetrad}, we get
\begin{equation}
     \mathcal{D}_\mu B^{ab}= h^{a}{ }_{\rho} h^{b}{ }_{\sigma}\Tilde{\nabla}_\mu B^{\rho \sigma} \ ,
\end{equation}
where $\Tilde{\nabla}_\mu B^{\rho \sigma}$ is the Levi-Civita covariant derivative of $B^{\rho \sigma}$ .\\
Thus, the Fock–Ivanenko derivative of the antisymmetric Lorentz tensor $B^{ab}$ reduces to
the usual Levi–Civita covariant derivative of general relativity. In other words, we can say that the minimal-coupling prescription in Riemannian geometry can be written as,
\begin{equation}
\partial_{a} \rightarrow \mathcal{D}_\mu =   \partial_{\mu}+\Tilde{\Gamma}_{\mu} \equiv \Tilde{\nabla}_{\mu} \ .
\end{equation}
Now let us consider the Kalb-Ramond Lagrangian in the background of Riemannian geometry.
\begin{equation}
    \mathcal{L}_m = -\sqrt{-g}H_{\mu \nu \rho}H^{\mu \nu \rho} \ ,
\end{equation}
where the field strength $H_{\mu \nu \rho}$ is given by,
\begin{equation}
    H_{\mu \nu \rho} = \Tilde{\nabla}_{\mu}B_{\nu \rho}+\Tilde{\nabla}_{\rho}B_{\mu \nu}+\Tilde{\nabla}_{\nu}B_{\rho \mu} \ .
\end{equation}
The corresponding field equation can be written as
\begin{equation}
    \Tilde{\nabla}_{\mu}H^{\mu \nu \rho} = 0 \ .
\end{equation}
Assuming the Lorentz gauge $\Tilde{\nabla}_{\mu}B^{\mu \nu}=0$, and using the commutation relation,
\begin{equation}
    \big[\Tilde{\nabla}_{\mu},\Tilde{\nabla}_{\nu}\big] B^{\lambda\mu}=    \Tilde{R}^{\lambda}{}_{\sigma\mu\nu}B^{\sigma\mu}+\Tilde{R}_{\mu \nu}B^{\lambda\mu} \ ,
\end{equation}
we have the field equations of KR fields in teleparallel geometry,
\begin{equation}
    \Tilde{\nabla}_{\mu}\Tilde{\nabla}^{\mu}B^{\nu \lambda} + \Tilde{R}^{\nu \lambda \sigma \mu}B_{\sigma\mu}+2\Tilde{R}_{\mu}{}^{[\nu}B^{\lambda]\mu} = 0 \ .
\end{equation}


\end{document}